\documentclass[aps,prb,twocolumn,floatfix,superscriptaddress,nofootinbib]{revtex4-2}
\usepackage{graphicx}
\usepackage{bbm}
\usepackage{amsmath}
\usepackage{amsfonts}
\usepackage{lipsum}
\usepackage[colorlinks=true,allcolors=blue]{hyperref}
\usepackage{ulem}
\usepackage{slashed}
\usepackage{tikz}
\usepackage[compat=1.1.0]{tikz-feynman}
\usetikzlibrary{decorations.markings}

\usepackage{comment}
\usepackage{subfigure}
\usepackage{booktabs}

\usepackage{braket} 
\usepackage{tensor} 

\newcommand{\mycomment}[1]{}

\newcommand{\nn}{\nonumber}
\newcommand{\un}{\underline}
\newcommand{\bb}{\mathbb}

\newcommand{\p}{\partial}
\newcommand{\tr}{\text{tr}}

\newcommand{\sgn}{\text{sign}}

\def\emph{\textit}

\begin{document}

\begin{abstract}
This work focuses on the non-dissipative, parity-odd spin transport of $(2+1)$-dimensional relativistic electrons, generated by torsion, and the torsional Hall viscosity $\zeta_{\rm H}$. We first determine $\zeta_{\rm H}$ for massive Dirac fermions in the presence of a constant electromagnetic field. We predict that the magnetic field induces a contribution to $\zeta_{\rm H}$ competing with the one originating from the Dirac mass. Moreover, we quantify the impact on $\zeta_{\rm H}$ originating from the band structure deformation quadratic in momentum terms that was proposed by Bernevig-Hughes-Zhang (BHZ). We find that the BHZ deformation substantially enhances $\zeta_{\rm H}$ in magnitude as measured in a domain wall configuration, when compared to the free Dirac fermion result. Nevertheless, the torsional Hall viscosity still discriminates between topologically trivial and non-trivial regimes. Our results, hence, pave the way for a deeper understanding of hydrodynamic spin transport and its possible verification in experiments.

\end{abstract}

\title{Torsional Hall Viscosity of Massive Chern Insulators: \\ Magnetic Field and Momentum Deformations}

\author{Ioannis~Matthaiakakis}
\email{I.Matthaiakakis@soton.ac.uk}
\affiliation{Mathematical Sciences and STAG Research Centre, University of Southampton, Highfield,
Southampton SO17 1BJ, United Kingdom}
\thanks{The first two authors contributed equally to this work.}

\author{Weizhen Jia}
\email{weizhenjia@cuhk.edu.hk}
\affiliation{Department of Physics, The Chinese University of Hong Kong, Shatin, New Territories, Hong Kong, China}
\thanks{The first two authors contributed equally to this work.}
\affiliation{Institute for Advanced Study, Tsinghua University, Beijing, 100084, China}
\affiliation{Institute for Theoretical Physics and Astrophysics, and W{\"u}rzburg-Dresden Cluster of Excellence on Complexity and Topology in Quantum Matter ct.qmat, Julius-Maximilians-Universit{\"a}t W{\"u}rzburg, Am~Hubland, D-97074~W{\"u}rzburg, Germany}

\author{Raffael~L.~Klees}
\affiliation{Institute of Physics, University of Augsburg, D-86159 Augsburg, Germany}

\author{David Rodr\'iguez Fern\'andez}
\affiliation{Departamento de matemática aplicada a las TIC, Universidad Politécnica de Madrid, Nikola Tesla, s/n, ES-28031 Madrid, Spain}

\author{Zhuo-Yu~Xian}
\affiliation{Department of Physics, Freie Universit\"at Berlin, Arnimallee 14, DE-14195 Berlin, Germany}

\author{Ren\'e~Meyer}
\affiliation{Institute for Theoretical Physics and Astrophysics, and W{\"u}rzburg-Dresden Cluster of Excellence on Complexity and Topology in Quantum Matter ct.qmat, Julius-Maximilians-Universit{\"a}t W{\"u}rzburg, Am~Hubland, D-97074~W{\"u}rzburg, Germany}

\author{Johanna~Erdmenger}
\affiliation{Institute for Theoretical Physics and Astrophysics, and W{\"u}rzburg-Dresden Cluster of Excellence on Complexity and Topology in Quantum Matter ct.qmat, Julius-Maximilians-Universit{\"a}t W{\"u}rzburg, Am~Hubland, D-97074~W{\"u}rzburg, Germany}

\author{Ewelina~M.~Hankiewicz}
\email{Ewelina.Hankiewicz@physik.uni-wuerzburg.de}
\affiliation{Institute for Theoretical Physics and Astrophysics, and W{\"u}rzburg-Dresden Cluster of Excellence on Complexity and Topology in Quantum Matter ct.qmat, Julius-Maximilians-Universit{\"a}t W{\"u}rzburg, Am~Hubland, D-97074~W{\"u}rzburg, Germany}
\affiliation{Department of Physics, University of California, Berkeley, CA 94720, USA}
\maketitle


\section{Introduction}\label{Sec:Intro}

 In the past two decades, we have seen the proliferation of distinct quantum field theoretic models as a means to describe and understand the low-energy physics of condensed matter systems (see, e.g., the reviews in Refs.~\cite{Wehling02012014,RevModPhys.90.015001,Chernodub_2022,Bradlyn:2014wla,Lucas_2018,Ong:2020ffe,Miransky:2015ava}). Namely, the linear transport and hydrodynamic regimes described by such models have been increasingly accessible in experimental settings and a variety of systems, allowing for the discovery and exploration of a wide range of physical phenomena \cite{Wehling02012014,Chernodub_2022,Bradlyn:2014wla,Lucas_2018,Ong:2020ffe}. Cataloging and understanding these phenomena is performed in terms of the system's transport coefficients, which dictate how the system distributes its conserved charges, such as energy and momentum, across space and time.

In this work, we consider spin transport in condensed-matter systems that exhibit an emergent relativistic invariance at low energies. Spin provides an additional mode of transport, along with energy and momentum, which is useful for both theoretical and experimental investigations \cite{PhysRevLett.100.236601,Ko_nig_2007,Bernevig_2006,Lu_2010,Hoyos_2014,HIROHATA2020166711}. The interplay between spin and torsion leads to nontrivial phenomena, in the sense that a torsionful background geometry couples to the spin of Dirac fermions, thus allowing us to probe nontrivial effects in the spin current. In addition, torsion appears naturally in lattice systems as the continuous limit of local lattice dislocation \cite{KATANAEV19921,KleinertGF,doi:10.1142/6742,PhysRevLett.106.161102,DEJUAN2010625}, and couples equally naturally to relativistic fermions \cite{Freedman2012,Hoyos_2014,Watanabe2004}. 

We focus on one particular spin transport coefficient, the torsional Hall viscosity $\zeta_{\rm H}$. The torsional Hall viscosity is a non-diffusive transport coefficient, present at both zero and finite temperatures, that is non-vanishing only in systems with broken parity symmetry. 
Its properties can be traced down to the topological properties of the material, such as its Berry curvature under torsionful deformations of the Hamiltonian \cite{Avron:1995fg,Hughes2011,Hughes2013}. It is also closely related to the parity torsional anomaly in $2+1$ dimensions, as first observed in the series of papers in Refs.~\cite{Hughes2011,Hughes2013,Parrikar:2014usa} (see also the review in Ref.~\cite{Hoyos_2014}). 

In earlier work on $\zeta_{\rm H}$, parity is explicitly broken and $\zeta_{\rm H}$ is induced by the fermion mass gap, i.e., the Dirac mass. In this work, we take a first step towards understanding the interplay of torsion with additional external sources. First, we focus on the interplay between torsion and constant external electromagnetic fields. Second, we deform the relativistic massive Dirac Lagrangian with a term quadratic in momentum, corresponding to the effective mass in 2D topological insulators. We refer to this term as the Bernevig-Hughes-Zhang (BHZ) deformation \cite{Bernevig_2006}. The inclusion of electromagnetic fields gives access to  transport phenomena such as the Shubnikov-de Haas effect \cite{Shubnikov} and more general Landau-level physics \cite{vonKlitzing:1980pdk,Thouless:1982zz}. The BHZ deformation, on the other hand, has arisen lately in the low-energy description of several experimentally available 2D topological materials, such as HgTe quantum wells \cite{Ko_nig_2007}, InAs/GaSb quantum wells \cite{PhysRevLett.100.236601,PhysRevB.81.201301} as well as BiSeTe thin films \cite{BiSeTe}. While the full BHZ model preserves time-reversal symmetry, a single spin block of the BHZ model breaks both time-reversal and parity symmetries, and therefore exhibits a nonzero torsional Hall viscosity. For example, materials described by half of the BHZ model are (Hg,Mn)Te quantum wells \cite{PhysRevLett.100.236601}. These are quantum anomalous Hall (QAH) insulators, since they have only one chiral edge state flowing at zero chemical potential.

The BHZ deformation breaks the effective Lorentz invariance of a massive Dirac system. As such, it is expected to arise naturally in condensed matter systems, which are intrinsically non-relativistic. We should note that both deformations considered in this paper have already been shown to affect charge transport and the Hall response  \cite{Bottcher:2019rrz,Tutschku:2020drw}. However, no such analysis exists for $\zeta_{\rm H}$. As a central result, we determine the impact of electromagnetic and BHZ deformations on the torsional Hall viscosity and spin transport. Note that in this work we consider the torsional Hall viscosity for massive Chern insulators at zero chemical potential and temperature. Our calculations are performed without any approximation on the magnitude of the electromagnetic field strength nor the BHZ deformation parameter. Therefore, our discussion applies to a physical system that is explicitly non-perturbative in both the electromagnetic field and the BHZ deformation parameter.  In this way, we bring the torsional transport closer to realistic lab conditions and enable its verification and fine-tuning by experimentalists. 

For the magnetic field deformation, we find that $\zeta_{\rm H}$ vanishes at a nonzero magnetic field strength, set by the Dirac mass gap, of ${\cal O}(100\,{\rm mT})$. In addition, we find that the leading order correction in the magnetic field strength $B$ (compared to the mass gap) scales as $B^2$ with a negative coefficient, providing what we call a negative ``magnetoviscosity". This is the first main result of this paper. We also examine the case of a constant electric field, in which we find a weak dependence of $\zeta_{\rm H}$ on perturbative electric field strengths. However, in the non-perturbative electric field strength regime, as relevant for non-perturbative quantum electrodynamics in materials (see, e.g., Ref.~\cite{Amoretti:2023dgb}), $\zeta_{\rm H}$ increases with the electric field and acquires a nontrivial imaginary part. This imaginary part arises because the electric field forces the system to tunnel to states with an ever-increasing number of excited electrons and holes, thus breaking unitarity.

For the BHZ deformation we find distinct phases for the torsional Hall viscosity, depending on the relative sign of the effective non-relativistic/Newtonian mass (represented by the BHZ deformation parameter) and the relativistic/Dirac mass gap, similar to the behavior of the corresponding BHZ Hall conductivity \cite{Tutschku:2020drw}. The torsional Hall viscosity thus shows a clear signature of a topological phase transition by taking different values in the topologically trivial and nontrivial regimes. We also note that the BHZ deformation can modify both the magnitude and sign of the torsional Hall viscosity compared to the standard massive Dirac fermion result. Unlike bulk and shear viscosities, which are dissipative and strictly non-negative, a negative Hall viscosity signifies that the perpendicular momentum current flows in the opposite direction to the Hall viscous force (see also Ref.~\cite{Hughes2013}). This is analogous to how a negative Hall conductivity indicates a flipped chirality of the edge modes in relation to the direction of the magnetic field. For typical (Hg,Mn)Te quantum wells, we further find that, for a domain wall across which the Dirac mass changes sign, the universal jump of the torsional Hall viscosity is enhanced by a factor of about $4.3$ compared to the pure Dirac case. This universal domain-wall response constitutes the second main result of this paper.

The paper is divided into the following sections. In Section~\ref{Sec:Setup}, we introduce the definition of the torsional Hall viscosity and summarize known results for the massive Dirac case to set the scene. In Section~\ref{sec:MagField}, we focus on a constant electromagnetic field and its effect on the torsional Hall viscosity. We continue in Section~\ref{sec:BHZ} with the BHZ deformation of the Dirac Lagrangian. Section \ref{sec:Conclusions} summarizes our main results and discusses possible research outlook.

\section{Setup and review of massive Dirac results}\label{Sec:Setup}

In this section, we first present our setup for the calculations appearing in Sections \ref{sec:MagField} and \ref{sec:BHZ}, and review the results for the torsional Hall viscosity of the free massive Dirac fermion \cite{Hughes2011,Hughes2013,Parrikar:2014usa}.

We begin by presenting our theoretical framework.
We introduce our background geometry and state the relation between the torsional Hall viscosity $\zeta_{\rm H}$ and the two-point function of Dirac fermions, see Eq.~\eqref{Eq:ZetaDeformedGreen} [and Eq.~\eqref{ZetaTadpole} for the corresponding Feynman diagram]. A more detailed description of the following setup can be found in Ref.~\cite{Hughes2013}.

We work in  a $(2+1)$-dimensional spacetime equipped with a metric $g_{\mu\nu}$ and an \emph{independent} affine connection $\nabla$ with coefficients $\tensor{\Gamma}{^\lambda_\mu_\nu}$ in a coordinate basis. 
We assume that the connection $\nabla$ is metric compatible, i.e., $\nabla_\mu g_{\rho\nu} = 0$. 
Given a general connection with no symmetries under exchange of its indices, the torsion tensor $\tensor{T}{^\lambda_\mu_\nu}$ of the background geometry is defined by its antisymmetric part $\tensor{T}{^\lambda_\mu_\nu} \equiv \tensor{\Gamma}{^\lambda_\nu_\mu} - \tensor{\Gamma}{^\lambda_\mu_\nu}$. Through a torsionful background geometry, we gain access to the torsional Hall viscosity of Dirac fermions \cite{Hoyos_2014,Hughes2011,Hughes2013}, as we demonstrate below for completeness.

Let us consider the coupling of a Dirac fermion $\psi$ to our background geometry. 
To do so, we introduce a dreibein frame field $\tensor{e}{^a_\mu}$ with $a,\mu = 0,1,2$.\footnote{We use Greek indices for holonomic, coordinate indices and Latin indices for anholonomic frame indices. Greek indices are raised and lowered with the coordinate metric $g_{\mu\nu}$, while Latin indices are raised and lowered with the Minkowski metric $\eta_{ab}$.}
The dreibein defines a local inertial frame, which pulls back the spacetime metric $g_{\mu\nu}$ to the Minkowski metric $\eta_{ab} = (-1,1,1)$ via $g_{\mu\nu} = \tensor{e}{^a_\mu}\tensor{e}{^b_\nu} \eta_{ab}$. 
We also define the Dirac matrices $\gamma^a$, obeying the Clifford algebra $\{\gamma^a,\gamma^b\}=2\eta^{ab}\mathbbm{1}$, as well as the spin connection
\begin{equation}
    \label{Eq:SpinToSpaceConn}
    \tensor{\omega}{^a_b_\mu} 
    \equiv \tensor{e}{^a_\lambda} \nabla_\mu \tensor{e}{_b^\lambda} 
    = \tensor{e}{^a_\lambda} (\partial_\mu \tensor{e}{_b^\lambda}  + \tensor{\Gamma}{^\lambda_\nu_\mu} \tensor{e}{_b^\nu} ) ,
\end{equation}
where $\tensor{\omega}{_a_b_\mu} + \tensor{\omega}{_b_a_\mu} = 0$ due to the assumed metric compatibility. 
Via the spin connection, we can define the covariant derivative of the Dirac spinor $\psi$ and its conjugate spinor $\bar{\psi} \equiv \psi^\dagger i\gamma^0$ as \cite{Freedman2012,Nakahara2003}
\begin{subequations}
\label{Eq:DiracCovariant}
\begin{align}
    \nabla_\mu \psi &= \partial_\mu\psi + \frac{1}{8}\omega_{ab\mu}[\gamma^a,\gamma^b] \psi,
    \\
    \nabla_\mu \bar{\psi} &= \partial_\mu\bar{\psi} - \frac{1}{8}\omega_{ab\mu} \bar{\psi} [\gamma^a,\gamma^b].
\end{align}
\end{subequations}
We follow the conventions of Ref.~\cite{Hughes2013} and define the Lorentzian version of the Dirac fermion action as
\begin{align}
    \label{Eq:DiracActionEC}
    S_\psi 
    &= \int  d^3x \, e \left( \frac{1}{2} \left[ 
     \bar{\psi} \gamma^\mu \nabla_\mu \psi 
     -
     \nabla_\mu \bar{\psi} \gamma^\mu \psi 
     \right]  
    - m \bar{\psi} \psi 
     \right) ,
\end{align}
where $\gamma^\mu \equiv \tensor{e}{_a^\mu} \gamma^a$ and $e \equiv |\mathrm{det}(\tensor{e}{^a_\mu})|$.

{{The action $S_\psi$ provides an effective description of the excitations in a Dirac material around the Dirac point. The spin degrees of freedom of $\psi$ correspond to an emergent pseudo-spin degree of freedom of the excitations due to the multi-valley nature of the Dirac material lattice.}} The action $S_\psi$ is not the most general action considered in this paper, but it suffices to demonstrate our formalism. The first step in evaluating the torsional Hall viscosity is to make explicit the coupling of $\psi$ to torsion. 
To do so, we separate the torsion contribution from the affine connection as $\tensor{\Gamma}{_\lambda_\mu_\nu} = \tensor{\mathring{\Gamma}}{_\lambda_\mu_\nu} + \tensor{K}{_\lambda_\mu_\nu}$, where $\tensor{\mathring{\Gamma}}{_\lambda_\mu_\nu} = (\partial_\mu g_{\nu\lambda} + \partial_\nu g_{\mu\lambda} - \partial_\lambda g_{\mu\nu}) / 2$ is the torsion-free Levi-Civita connection and
\begin{align}
    \tensor{K}{_\lambda_\mu_\nu}
    =
    \frac{1}{2} (T_{\mu\lambda\nu} - T_{\lambda\mu\nu} + T_{\nu\lambda\mu})
\end{align}
is the contorsion tensor, which carries all torsion information of the background geometry. It is antisymmetric in the first and second indices, $\tensor{K}{_\lambda_\mu_\nu} + \tensor{K}{_\mu_\lambda_\nu} = 0$, and torsion can be expressed explicitly in terms of the contorsion tensor as $T_{\lambda\mu\nu} = \tensor{K}{_\lambda_\nu_\mu} - \tensor{K}{_\lambda_\mu_\nu}$.
The split of the spacetime connection into Levi-Civita and contorsion pieces translates to a similar split of the spin connection $\tensor{\omega}{^a_b_\mu} = \tensor{\mathring{\omega}}{^a_b_\mu} + 
    \tensor{e}{^a_\lambda} \tensor{K}{^\lambda_\nu_\mu} \tensor{e}{_b^\nu}$, 
with the Levi-Civita spin connection $\tensor{\mathring{\omega}}{^a_b_\mu} 
    \equiv \tensor{e}{^a_\lambda} (\partial_\mu \tensor{e}{_b^\lambda}  + \tensor{\mathring{\Gamma}}{^\lambda_\nu_\mu} \tensor{e}{_b^\nu} )$.

We can separate the contorsion, and hence the torsion, contribution from the action as $S_\psi = \mathring{S}_\psi + S_K$. The action $\mathring{S}_\psi$ is defined as $S_\psi$ in Eq.~\eqref{Eq:DiracActionEC}, but with $\nabla_\mu$ replaced by the torsion-free connection in terms of the Levi-Civita spin connection $\tensor{\mathring{\omega}}{^a_b_\mu}$. The action $S_{K}$ is in turn defined as 
\begin{align}
    S_K 
    &= \int  d^3x \, e \, \frac{1}{16} K_{abc} \bar{\psi} \{  [\gamma^a , \gamma^b] , \gamma^c \} \psi ,
    \nn \\
    &= \int  d^3x \, e \, \frac{1}{8} K_{abc} \bar{\psi} (\gamma^b \gamma^c \gamma^a -\gamma^a \gamma^c \gamma^b  ) \psi ,
\end{align}
with $K_{abc} \equiv \tensor{e}{_a^\lambda} \tensor{e}{_b^\mu} \tensor{e}{_c^\nu} K_{\lambda\mu\nu}$.
Note that the term in brackets is completely antisymmetric, i.e., it is zero if two or more indices take the same value.
This means that only the completely antisymmetric part of the contorsion tensor, $K_{[abc]}$, contributes to the action $S_K$, which allows us to rewrite it as\footnote{The antisymmetrization of $p$ indices includes the factor $1/p!$, e.g., $A_{[ab]} = (A_{ab}- A_{ba})/2.$}
\begin{align}
    S_K 
    &= \int  d^3x \, e \, \frac{1}{8} K_{[abc]} \bar{\psi} (\gamma^{[b} \gamma^c \gamma^{a]} -\gamma^{[a} \gamma^c \gamma^{b]}  ) \psi
    \nn \\
    &= \int  d^3x \, e \, \frac{1}{4} K_{[abc]} \bar{\psi} \gamma^{[a} \gamma^b \gamma^{c]}  \psi.
\end{align}
This proves again the well-known fact that Dirac fermions couple only to the completely antisymmetric part of torsion \cite{Shapiro_2002}. Our calculation so far is valid in any spacetime dimension. Upon focusing only on $2+1$ dimensions, we may use the Dirac matrix identity, $\gamma^{[a} \gamma^b \gamma^{c]} = \varepsilon^{abc} \mathbbm{1}$.\footnote{The sign of $\gamma^{[a} \gamma^b \gamma^{c]} = \pm \varepsilon^{abc} \mathbbm{1}$ is a choice of orientation, which is reflected in the sign of $\zeta_{\rm H}$. We choose the positive sign to match the choice in \cite{Hughes2013}.} We can therefore define the torsion pseudoscalar $\sigma =  \, \varepsilon^{abc} K_{[abc]} = \varepsilon^{\lambda\mu\nu} K_{[\lambda\mu\nu]}$, such that 
\begin{align} \label{eq:torsionPseudoscalarDefinition}
    S_K 
    = \int  d^3x\, e \, \frac{1}{4}  \bar{\psi} \sigma \psi.
\end{align}

The torsion pseudoscalar $\sigma$, through $S_K$, gives rise to the torsional Hall viscosity $\zeta_{\rm H}$, dictating the response of $\psi$ to deformations of the dreibein. This response is captured by the effective action \cite{Hoyos_2014}
\begin{equation}
\label{Eq:ZetaEfAction}
    S_\mathrm{eff} = \frac{\zeta_{\rm H}}{2} \int d^3x\,e\, \eta_{ab}\varepsilon^{\lambda\mu\nu}\tensor{e}{^a_\lambda} {\overset{\omega}{\nabla}}_\mu \tensor{e}{^b_\nu} 
    = 
    -{ \zeta_{{\rm H}}\over 2}\int d^3x\,e\,\sigma ,
\end{equation}
where ${\overset{\omega}{\nabla}}_\mu \tensor{e}{^b_\nu} = \partial_\mu \tensor{e}{^b_\nu} + \tensor{\omega}{^b_c_\mu} \tensor{e}{^c_\nu}$.
 The effective action $S_{\rm eff}$ depends on both the dreibein and the spin connection. Hence, $\zeta_{\rm H}$ contributes to both momentum and spin transport in the system. For this reason, in the sequel we discuss the effects of $\zeta_{\rm H}$ on momentum and spin currents interchangeably. {{We also want to note that while $S_{\rm eff}$ is a ``Chern-Simons term'' for the dreibein, we cannot infer that $\zeta_{\rm H}$ is quantized. This is because $\tensor{e}{^a_\mu}$ is a well-defined field in spacetime, unlike e.g. a $U(1)$ gauge field. Consequently, we cannot invoke any form of gauge principle to quantize the value of $\zeta_{\rm H}$.} }

We review how to arrive at $S_{\rm eff}$ directly from $S_\psi$ in Appendix~\ref{appendix:pathIntegralZetaH}. In this way, we obtain
\begin{equation}
    \label{Eq:ZetaGreen}
    \zeta_{\rm H} = \frac{1}{2} \lim_{y\to x}\tr[ G_{0}(x;y) ], 
\end{equation}
where $\tr(\, \cdot\, )$ is the trace over spin degrees of freedom and $G_{0}$ is the Dirac fermion propagator in a torsion-free background, derived from $\mathring{S}_\psi$.
Diagrammatically, the torsional Hall viscosity for the torsion-free Dirac fermion is given by the following tadpole diagram
\begin{align}
\label{ZetaTadpole0}
\zeta_{\rm H} = {1 \over 2}\feynmandiagram [inline=(b.base),layered layout, horizontal=a to b] {
a [crossed dot, label=\(\sigma\)]-- [scalar] b-- [fermion, half left, looseness=1.6,edge label=\(\)] c,
c-- [fermion, half left, looseness=1.6,edge label=\(\)] b};~.
\end{align}

In a similar fashion, we can derive the torsional Hall viscosity for systems perturbed away from the free Dirac-fermion case. 
In general, both the propagator $G_{0}$ and the vertex may be modified. 
Here, we are only interested in the part of the deformation linear in $\sigma$, which we assume is quadratic in $\psi$. That is, we assume that $S_\psi$ can be expressed as
\begin{equation}
    \label{Eq:DeformedS}
    S_\psi = \int d^3x\, e \left( \bar{\psi}G^{-1}\psi + \bar{\psi}V\sigma\psi\right)~,
\end{equation}
where $G$ is the torsion-free Dirac propagator, possibly in the presence of additional external fields, while $V$ is the vertex function between torsion and $\psi$.
 
In this case, $\zeta_{\rm H}$ is given by (see Appendix~\ref{appendix:pathIntegralZetaH})
\begin{equation}
    \label{Eq:ZetaDeformedGreen}
    \zeta_{\rm H} = 2V\lim_{y \rightarrow x} {\rm tr}[G(x;y)],
\end{equation}
where $G$ is again the torsion-free propagator, containing possible couplings to other fields or higher-derivative fermion terms. In turn, $V$ is the vertex function defined by a given deformation. For the torsion-free Dirac fermion we have $G=G_{0}$ and $V = 1/4$, as can be read off from Eq.~\eqref{eq:torsionPseudoscalarDefinition}.
Equation \eqref{Eq:ZetaDeformedGreen} can be represented by a dressed tadpole diagram:
\begin{align}
\label{ZetaTadpole}
\zeta_{\rm H} = 2\feynmandiagram [inline=(b.base),layered layout, horizontal=a to b] {
a [crossed dot, label=\(\sigma\)]-- [scalar] b [label=\(V~~\)] -- [fermion, half left, looseness=1.6,edge label=\(\)] c,
b-- [photon, half left, looseness=1.6,edge label=\(\)] c,
c-- [photon, half left, looseness=1.6,edge label=\(\)] b,
c-- [fermion, half left, looseness=1.6,edge label=\(\)] b};~,
\end{align}
where the wiggly line denotes the propagator $G$, possibly dressed by other sources. 

To compute $\zeta_{\rm H}$ by evaluating $\tr [G_{0}]$, we need to regularize the divergences that show up in vacuum bubble diagrams. In this paper, we employ either a heat kernel/$\zeta$-function regulator \cite{Vassilevich:2003xt} or a Pauli-Villars regulator. In the rest of this section, we present the case of massive fermions, as discussed in detail in Ref.~\cite{Hughes2013}, to exemplify the heat-kernel method, which is also applied in Section~\ref{sec:MagField} for the case of a constant electromagnetic field. An explicit discussion of the Pauli-Villars method is left for Section~\ref{sec:BHZ}, where we consider the BHZ model.

Consider a flat background geometry with $g_{\mu\nu} = \eta_{\mu\nu}$ and $\omega_{ab\mu} = 0$.
Defining the operators $\slashed{\partial} = \gamma^\mu \partial_\mu$, the propagator of the torsion-free Dirac fermion simply reads
\begin{equation}
\label{G0}
\begin{aligned}
G_{0} = {1 \over \slashed{\partial} - m } =  -{\slashed{\partial}+ m  \over -\partial^2 + m^2}.
 \end{aligned}
\end{equation}
{{Upon taking the trace to evaluate $\zeta_{\rm H}$, we make use of $\tr(\gamma^\mu) = 0$. Hence, we may drop the first term in the numerator on the right-hand side of the last equality and, without loss of generality, use the propagator in the form}}
\begin{equation}
    G_0 \sim  {-m \over -\partial^2 + m^2} = -m \int_0^\infty ds \, K(s) . 
\end{equation}
We have now also introduced the heat kernel of the square of the Dirac operator:
\begin{align}
K(s) \equiv \exp[-s(-\partial^2 + m^2)]\,.
\end{align}
To evaluate the propagator at coincident spacetime points, we can either take the Fourier transform of $(-\partial^2 +m^2)^{-1}$ or diagonalize the heat kernel. In the latter case, the heat-kernel eigenfunctions are plane wave spinors 
\begin{equation}
    \chi_{s,\Vec{k}} = {1\over (2\pi)^{3/2}}u_s e^{-ik\cdot x}~~,~~s =\uparrow, \downarrow~,
    \label{Eq:FreeWaves}
\end{equation}
with $u_s^\dagger u_s =1$. Then, using $V(y) = 1/4$, $\zeta_{\rm H}$ in terms of the heat kernel becomes
\begin{align}
    \label{Eq:VacuumZeta1}
    \zeta_{\rm H} &= -{m\over 2}\int_0^\infty \frac{d^3k}{(2\pi)^3} {1 \over k^2+m^2}\nn\\
    &= -{m\over 2}\int_0^\infty ds\int {d^3k \over (2\pi)^3}e^{-(k^2+m^2)s} \nn
    \\
    &=-{\rm sign(m)}{m^2\over 4\pi} + {m\Lambda \over 2\pi^2} + {\cal O}(\Lambda^{-1})\,,
\end{align}
where we introduced a cutoff $\Lambda$. After properly renormalizing, the universal part\footnote{There is also a non-universal term appearing on both $m>0$ and $m<0$, which depends on the renormalization scheme. However, the difference of $\zeta_{\rm H}$ in the two phases is universal.} of the torsional Hall viscosity reads \cite{Hughes2013}
\begin{align}
\label{Eq:ZetaHLP}
2\pi\zeta_{\rm H}=
\begin{cases}
0\quad& m>0\,,\\
m^2\quad& m<0\,.
\end{cases}
\end{align}
We can recognize the case with $m>0$ as the trivial phase and $m<0$ as the topological phase, similar to the picture of the Hall conductivity $\sigma_{\rm H}$ \cite{Tutschku:2020drw}.

\section{Turning on a magnetic field}\label{sec:MagField}

In this section, we consider the first of the two deformations of $S_\psi$ that we study in this paper. Namely, we turn on a constant magnetic field $B$, perpendicular to the two-dimensional system, and evaluate the torsional Hall viscosity in the resulting vacuum. We also show how to use the constant magnetic field result in order to obtain $\zeta_{\rm H}$ in the presence of a constant electric field. 
 
We consider again a flat background geometry with $g_{\mu\nu} = \eta_{\mu\nu}$ and $\omega_{ab\mu}=0$, and introduce a background $U(1)$ gauge field $A_\mu$. 
In this case, the $\psi$ propagator reads $G = (\slashed{D}-m)^{-1}$ , where
\begin{equation}
\slashed{D} = \gamma^\mu (\partial_\mu -ie A_\mu)
\end{equation}
and $e$ is the electron charge.
To evaluate the torsional Hall viscosity, we use the heat kernel approach. As a first step, we write
\begin{align}
    \label{Eq:QuadProp}
    \tr [G] 
    &= {\rm tr}\left[{1 \over \slashed{D} -m}\right] 
    = {\rm tr}\left[{(-\slashed{D}-m) \over (\slashed{D} -m)(-\slashed{D}- m)}\right] 
    \nonumber
    \\
    &=  -m\, {\rm tr}\left[{1 \over -\slashed{D}^2 +m^2}\right] -{\rm tr}\left[{\slashed{D} \over -\slashed{D}^2 +m^2}\right]
    \nonumber
    \\
    &=-m\, {\rm tr}\left[{1 \over -\slashed{D}^2 +m^2}\right]
    \equiv -m\, {\rm tr}[H^{-1}] ~.   
\end{align}
In the last line, we dropped one of the terms, as it vanishes due to the fermion trace\footnote{This is only true because we assume that the Maxwell field strength $F_{\mu\nu} = \partial_\mu A_\nu - \partial_\nu A_\mu$ has only one nontrivial component given by the magnetic field $B = F_{xy} = - F_{yx}$.} and defined the operator $H = m^2 - \slashed{D}^2$ for later convenience. Therefore, we may write 
\begin{equation}
    \label{Eq:ZetaHeatKernel}
    \begin{aligned}
   \zeta_{\rm H} &= -{m\over 2}\lim_{y \rightarrow x}\int_0^\infty ds ~{\rm tr} \langle x| e^{-sH}|y\rangle 
   \\
   &= -{m\over 2}\lim_{y \rightarrow x}\int_0^\infty ds~{\rm tr}[K(s;H)]~,
   \end{aligned}
\end{equation}
where we have introduced the heat kernel $K(s;H)$.
Note that $\zeta_{\rm H}$ is a functional of the background field $A_\mu$. By expanding around $A_\mu = 0$, one can consider Eq.~\eqref{Eq:ZetaHeatKernel} as computing the following dressed tadpole diagram by resumming the 1-loop diagrams linear in $\sigma$ but with insertions of all powers of $A_\mu$:
\begin{align}
\label{dressedloop}
\feynmandiagram [inline=(b.base),layered layout, horizontal=a to b] {
a [crossed dot, label=\(\sigma\)]-- [scalar] b-- [fermion, half left, looseness=1.6] c,
b-- [photon, half left, looseness=1.6] c,
c-- [photon, half left, looseness=1.6] b,
c-- [fermion, half left, looseness=1.6] b
};
~& =~  
\feynmandiagram [inline=(b.base),layered layout, horizontal=a to b] {
a [crossed dot, label=\(\sigma\)]-- [scalar] b-- [fermion, half left, looseness=1.6] c
-- [fermion, half left, looseness=1.6] b};
\\
&+~ 
\begin{tikzpicture}[baseline=(sigma.base)]
  \begin{feynman}
    \vertex[circle, fill=white, draw=black, thin, inner sep=2pt, 
            minimum size=3pt, label=above:$\sigma$] (sigma) at (0,0) {};
    \vertex (b) at (1,0);
    \vertex (c) at (2,0);
    \vertex[circle, fill=white, draw=black, thin, inner sep=2pt, 
            minimum size=3pt, label=above:$A_\mu$] (A) at (3,0) {};
    \draw[thin, line width=0.4pt] 
      (sigma.center) +(-0.07,-0.07) -- +(0.08,0.08)
      (sigma.center) +(-0.07,0.07) -- +(0.08,-0.08);
    \draw[thin, line width=0.4pt] 
      (A.center) +(-0.07,-0.07) -- +(0.08,0.08)
      (A.center) +(-0.07,0.07) -- +(0.08,-0.08);
    \diagram* {
      (sigma) -- [scalar] (b)
              -- [fermion, half left, looseness=1.5] (c)
              -- [fermion, half left, looseness=1.5] (b),
      (c) -- [photon] (A),
    };
  \end{feynman}
\end{tikzpicture}
~+~\cdots \nonumber
\end{align}
To evaluate $\zeta_{\rm H}$, we first calculate the heat kernel of a Dirac fermion in an external electromagnetic field. We do this by expressing $K(s;H)$ in terms of the eigenfunctions of $H$. More precisely, suppose $\chi_\lambda$ are the eigenfunctions of $H$ with eigenvalues $\lambda$, we have\footnote{We assume $\lambda$ is discrete in Eq.~\eqref{Eq:ZetaEigenvalues}, but extending to a continuous spectrum is formally trivial.}
\begin{equation}
    \label{Eq:ZetaEigenvalues}
    \zeta_{\rm H} = -{m \over 2}\lim_{y \rightarrow x}\int_0^\infty ds \sum_\lambda e^{-\lambda s}{\rm tr}[\chi^\dagger_\lambda(x) \chi_\lambda(y)]~,
\end{equation}
where we may choose to perform the sum or integral over $\lambda$ before the integral over $s$ depending on the problem at hand.

To make our calculation precise, we Wick rotate time and work with a positive definite flat metric $g_{\mu\nu} = \delta_{\mu\nu}$. Indices now run over $1, 2,3$, with $\mu =3$ being the Wick-rotated time direction. In addition, we choose the Euclidean background gauge field $A_\mu = (A_3,A_1,A_2) = (0,0,B x^1)$, which represents a constant magnetic field pointing out of the plane. 
In this gauge, $H$ is the Hamiltonian of a harmonic oscillator, and finding its eigenfunctions is a textbook problem \cite{strange_1998} (see also Ref.~\cite{Miransky:2015ava}). For completeness, we present the solution in Appendix~\ref{App:VacuumSol}, and below we quote the final result for the wavefunctions
\begin{align}
    \label{Eq:SquareSolution'}
    &\psi_{r, n, p_2, p_3} = N u_r e^{ip_3x^3 + ip_2x^2}e^{-X^2/2}H_n(X)
    ~,
\end{align}
with eigenvalues $\lambda_r = ({2n+ 1 -{\rm sign}(eB)r)/ l_B^2} + p_3^2 +m^2$. The quantum numbers are the continuous momenta  $p_2$ and $p_3$, while $r=\pm$ and $n \geq 0$ are discrete spin and energy-level indices, respectively. 
We introduce the magnetic length $l_B^2 = 1/|eB|$ and denote the $n$th Hermite polynomial by $H_n$. The remaining definitions used are
\begin{align}
    X &= {x\over l_B} + p_2l_B\,,\qquad N^2 = {1\over (2\pi)^2\sqrt{\pi}2^n n! l_B}\,,
    \\
    \nonumber
    u_+ &= {1\over \sqrt{2}}\begin{pmatrix}
        1\\0\end{pmatrix}\,,\qquad u_-= {1\over \sqrt{2}}\begin{pmatrix}
            0\\1 \end{pmatrix}\,.
\end{align}
Using Eq.~\eqref{Eq:SquareSolution'}, we can evaluate the sum over $\lambda$ in Eq.~\eqref{Eq:ZetaEigenvalues} in closed form (see Appendix~\ref{App:VacuumSol}) to find
\begin{equation}
    \label{Eq:FiniteBHeat}
    {\rm tr}[K(s; H)] = {\sqrt{\pi}\over 4\pi^2l^3_B \sqrt{\tilde{s}}}e^{-\tilde{m}^2\tilde{s}}\coth({\tilde{s}})~.
\end{equation}
The parameters $\tilde{m}= ml_B$ and $\tilde{s} = s/l_B^2$ are  dimensionless versions of $m$ and $s$ respectively. To complete our calculation in Eq.~\eqref{Eq:ZetaHeatKernel}, we are thus left with calculating the integral of Eq.~\eqref{Eq:FiniteBHeat} over $\tilde{s}$,
\begin{equation}
\label{Eq:AuxZetaB}
    \zeta_{\rm H} = - {m\over 4\pi^{3/2} l_B} I(\tilde{m}^2)~,
\end{equation}
with 
\begin{equation}
\label{Eq:AuxIm}
    I(\tilde{m}^2) = \int_0^\infty ds {e^{-\tilde{m}^2s} \over \sqrt{s}}\coth({s})~.
\end{equation}
It is not surprising that this integral suffers from a divergence at $\tilde{s} = 0$, which is a remnant of the infinite momentum part of the spectrum. One can employ a cutoff to regulate the integral, as in the case of the vacuum result presented in Section~\ref{Sec:Setup}. We, however, employ $\zeta$-function regularization to make the integral finite. In particular, we have
\begin{equation}
\label{Eq:RegInt}
\begin{aligned}
I(\tilde{m}^2) &= {1\over 2\sqrt{2}}\int_0^\infty ds\left[{1\over \sqrt{s}}{e^{-\tilde{m}^2s/2}\over 1-e^{-s}} + {1\over \sqrt{s}}{e^{-(2+\tilde{m}^2)s/2}\over 1-e^{-s}}\right] 
\\
&\equiv {\sqrt{2\pi} \over 4}\left[2\zeta(1/2,\tilde{m}^2/2)- \sqrt{2\over \tilde{m}^2}\right],
\end{aligned}
\end{equation}
where in the last equality we used an integral representation of the Hurwitz $\zeta$-function
\begin{equation}
    \zeta(x,a) = {1\over \Gamma(x)}\int_0^\infty ds~s^{x-1} {e^{-as}\over 1-e^{-s}}~.
    \label{Eq:ZetaHDef}
\end{equation}
While the right-hand side of Eq.~\eqref{Eq:ZetaHDef} is valid for ${\rm Re}(x)>1$, the Hurwitz $\zeta$-function $\zeta(x,a)$ can be analytically continued to ${\rm Re}(x)<1$ and, in particular, to $x=1/2$. We use this analytic continuation as the regularization of $I(\tilde{m}^2)$, and hence the $\equiv$ sign in the final equality. 

Substituting $I(\tilde{m}^2)$ from Eq.~\eqref{Eq:RegInt} into Eq.~\eqref{Eq:AuxZetaB} for $\zeta_{\rm H}$, we find
\begin{align}
    \zeta_{\rm H} &= -{\sqrt{2}m \over 16\pi l_B}\left[2\zeta(1/2,\tilde{m}^2/2) -\sqrt{2\over \tilde{m}^2}\right] \nn
    \\
    &=-{\sqrt{2|eB|}m \over 16\pi}\left[2\zeta\left({1\over 2}, {m^2 \over 2|eB|}\right)-\sqrt{2|eB|\over m^2}\right] \nonumber
    \\
    \label{Eq:TorViscInB}
    &=-{\sgn}(m){m^2v_{\rm F}^4 \over 16\pi}\sqrt{2|B|\over B_c}\Bigg[2\zeta\bigg({1\over 2}, {B_c \over 2|B|}\bigg)-\sqrt{2|B|\over B_c}\Bigg]\,,
\end{align}
where in the second equality we re-introduced $eB$ and $m$ explicitly, and in the third equality we went back to SI units and introduced the Schwinger magnetic field limit $B_c = m^2v_{\rm F}^2/|e|\hbar$, which may be more familiar to some readers. 

\begin{figure}[!t]
\centering
    {\includegraphics[scale=0.9]{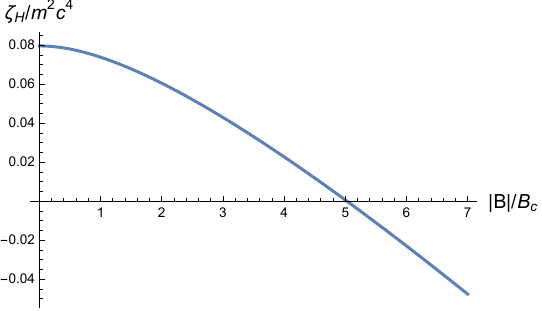}}
    \caption{$\zeta_{\rm H}$ in units of the electron rest mass as a function of the dimensionless ratio $B/B_c$, where $B_c$ is the Schwinger limit. In making this plot, we have assumed $m>0$.}
    \label{fig:zetavsB1}
 \end{figure}

\begin{figure}[!b]
\centering
    {\includegraphics[scale=0.9]{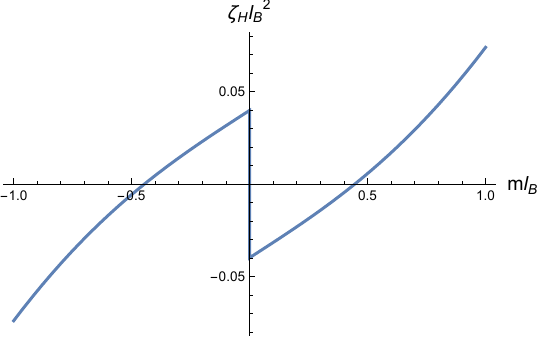}}
    \caption{$\zeta_{\rm H}$ in units of the magnetic length $l_B$ as a function of the dimensionless mass $\tilde{m}=ml_B$.}
    \label{fig:zetavsB2}
\end{figure}

We plot $\zeta_{\rm H}$ as a function of $B$ and $m$ in Figs.~\ref{fig:zetavsB1} and \ref{fig:zetavsB2}, respectively. We observe that $\zeta_{\rm H}$ is an odd function of the mass but an even function of $B$. This is a consequence of $\zeta_{\rm H}$ being a Lorentz pseudoscalar. Because of Lorentz invariance, $B$ may appear only through contractions of the Maxwell tensor $F_{\mu\nu}$, i.e., only even powers of $B$ are allowed in the expansion in Eq.~\eqref{dressedloop}. 
A resummation of this series can then only yield a function of $B^2$ (e.g., $\sqrt{B^2} = |B|$). 
Thus, we are left with the mass as the only parameter that can make $\zeta_{\rm H}$ a pseudoscalar; $m$ is parity odd and hence $\zeta_{\rm H}$ must change sign whenever $m$ does. We further observe that $\zeta_{\rm H}$ is a \emph{monotonically decreasing} function of $|B|$, but a monotonically increasing function of $m$ for $m>0$. The latter quality is present already in the case of a vanishing magnetic field \cite{Hughes2013}, and here we show that it persists for any nonzero $B$.

The monotonic decrease with respect to $|B|$ is one of the novel results of this paper, showcasing that Dirac fermions exhibit what we may call a negative torsional magnetoviscosity. The most surprising feature of this behavior is the nontrivial zero that appears for a nonzero value of $B$, numerically found to be located at $|B| \equiv B_0 = 5.02817 B_c$.\footnote{This behavior resembles the one found for the charge current in Fermi and Dirac liquids due to the regular Hall viscosity at perturbative magnetic field strengths \cite{Matthaiakakis:2019swi,Xian:2022eur}.}
Furthermore, $B=0$ appears to be a singular/indeterminate point for $\zeta_{\rm H}$. This fact, along with the square root dependence of the final result, indicates that the zero in $\zeta_{\rm H}$ is generated by non-perturbative effects in the magnetic field strength. 

We can explore these phenomena further by observing the behavior of $\zeta_{\rm H}$ in the limits $m\rightarrow 0 $ and $B\rightarrow 0$. We have\footnote{Note that the expansion around $B = 0$ is an asymptotic one (see, e.g., Ref.~\cite{Aniceto_2019}).}
\begin{widetext}
    \begin{align}       
    \label{Eq:ZetaZeroM}
        \zeta_{\rm H}(m\rightarrow 0, B) &= -{1\over 8\pi}{\rm sign}(m)|eB| -{\zeta(1/2)\over 4\sqrt{2}\pi}m\sqrt{|eB|} + {\cal O}\left(m^3\right)~,\\
\label{Eq:TorZetaPertB}
        \zeta_{\rm H}(m, B\rightarrow 0) &\sim {\rm sign(m)}{m^2 \over 4\pi}\left[1- {1 \over 4}\sum_{k=1}^\infty {(1/2)_{2k-1}B_{2k}\over (2k)!}\left({2eB \over m^2} \right)^{2k}\right]~,
    \end{align}
\end{widetext}
where $B_{n}$ are the Bernoulli numbers, and $(x)_n$ is the Pochhammer symbol. 
    
By examining the first few terms of Eq.~\eqref{Eq:TorZetaPertB}, we observe that the decrease in the value of $\zeta_{\rm H}$ with $B$ stems from a competition between the Dirac mass and magnetic field contributions. This competition has the same origin as the renormalization of the Hall conductivity discussed in Ref.~\cite{Bottcher:2019rrz} and can be traced to the spectral asymmetry of the Dirac operator and, effectively, to ``curing" the parity anomaly in gauge invariant systems {\cite{Redlich:1983dv}}. From Eq.~\eqref{Eq:TorZetaPertB} one can observe that the renormalization scheme here differs from that of Ref.~\cite{Hughes2013} as presented in Eq.~\eqref{Eq:ZetaHLP}. Our scheme treats the $m>0$ and $m<0$ phases equally, giving both a non-zero value for $\zeta_{\rm H}$. Of course, the difference between the two phases is universal and is given precisely by the result of Ref.~\cite{Hughes2013}, as expected.
    
We see that a nonzero value of $\zeta_{\rm H}$ persists even when the mass goes to zero. This is expected since $B$ is a parity-breaking coefficient and is, in principle, enough to generate a Hall response. Notice though that the Hall response generated solely by the magnetic field depends on how we take the mass to zero. This is because $B$ alone cannot make $\zeta_{\rm H}$ a pseudoscalar, as discussed earlier. This mass-dependence also indicates that we can generate domain walls with discontinuities in $\zeta_{\rm H}$. This is another hint that the torsional Hall viscosity is sensitive to the quantum anomalies of massless fermions.
    
Furthermore, the expansion around $B = 0$ in Eq.~\eqref{Eq:TorZetaPertB} shows that our main result in Eq.~\eqref{Eq:TorViscInB} for the torsional Hall viscosity is the resummation to all orders in $B$ of the perturbative series result, represented by the Feynman diagrams in Eq.~\eqref{dressedloop}. Note that, as is generally expected for perturbative series in QFT \cite{Dyson:1952tj}, our perturbative series is only asymptotic, hence the symbol $\sim$ is used. More precisely, the infinite sum in Eq.~\eqref{Eq:TorZetaPertB} diverges due to the factorial growth of the summand coefficients. However, since the series is alternating, it can be Borel-resummed to the $\zeta_{\rm H}$ given by Eq.~\eqref{Eq:TorViscInB}. 
   
We emphasize that the resummed series depends on the ratio ${eB/ m^2}$, which is closely related to the Schwinger limit of QED. This ratio becomes of order 1 when the magnetic field approaches the Schwinger limit, at which the vacuum becomes birefringent. For vacuum QED this limit is ${\cal O}(10^9) {\rm T}$ and goes far beyond any magnetic field in a lab. However, this is not the case for electrons in a solid. When confined to a solid, we may identify the mass of the electron with the gap $\Delta$ in the band structure, and write the critical magnetic field as $B_c = \Delta^2/(|e|\hbar v_{\rm F}^2)$. The Fermi velocity $v_{\rm F}$ must be calculated self-consistently from the electron-electron interactions, as shown, e.g., in Ref.~\cite{DiSante2020}. For a gap $\Delta ={\cal O} (1)$ meV and $ v_{\rm F} = {\cal O} (10^5)$ m/s, we find a critical magnetic field $B_c = {\cal O}(100~{\rm mT})$. This is well within the reach of current experimental probes and shows that our results for $\zeta_{\rm H}$ are in principle experimentally verifiable.

The fact that $\zeta_{\rm H}$ depends on $B$, or more generally the gauge field $A_\mu$, indicates that the effective action in $S_{\rm eff}$ in Eq.~\eqref{Eq:ZetaEfAction} does not give rise only to a torsional Hall viscosity, but also to a ``torsional Hall conductivity'' originating from higher-order corrections in $A_\mu$, pictorially represented in Eq.~\eqref{dressedloop}. This effect is similar to the case of a non-relativistic Hall viscosity $\eta_{\rm H}$ which stems from the Wen-Zee term \cite{Wen:1992ej}.\footnote{\label{FN:WZ}For non-relativistic  fermions, the effective theory has a Wen-Zee term $\eta_{\rm H}\int\epsilon^{\mu\nu\rho} \omega_\mu\p_\nu A_\rho$, where $\omega_\mu$ is the spin connection for the $2d$ spatial rotation $SO(2)$ \cite{Diamantini:1992uu,Abanov:2014ula}. This term is not only responsible for the Hall viscosity, but also gives rise to a correction to the Hall conductivity.} The coupling between the electromagnetic field and the spin connection modifies the charge density distribution in the presence of a nontrivial background geometry, which in turn leads to a higher-order correction to the Hall conductivity \cite{Hoyos:2011ez}.

\subsection{Rotating to a constant electric field}

In this subsection, we employ our results for $\zeta_{\rm H}$ as a function of a constant magnetic field, as shown in Eqs.~\eqref{Eq:TorViscInB} and \eqref{Eq:TorZetaPertB}, to compute $\zeta_{\rm H}$ as a function of a constant electric field. 

The dependence on the Schwinger ratio along with the emergent Lorentz invariance of our theory suggests the existence of a torsional Hall viscosity in a finite electric field $\Vec{E}$.\footnote{The electric field used here propagates in $2+1d$, so it differs from the electromagnetic field in an experiment. By using a $2+1d$ electric field, we are effectively neglecting the Coulomb interaction between electrons. This does not limit our results as we are working at charge neutrality.} In particular, Lorentz and $U(1)$ gauge invariance suggest that $\zeta_{\rm H}$ must be a function of the scalars constructed from the Maxwell field strength $F_{\mu\nu} = \partial_{\mu}A_{\nu} - \partial_{\nu}A_{\mu}$.\footnote{Our assumptions can be weakened to just gauge invariance, which enforces the minimal coupling of the Dirac fermions to the gauge field and the functional form of $\zeta_{\rm H}$}  The only scalars in $2+1$ dimensions are
\begin{equation}
    \label{Eq:LorentzScalar}
F^2 \equiv F^{\mu\nu}F_{\mu\nu} = B^2 - \vec{E}^2,
\end{equation}
and powers of it. Therefore, for a constant electromagnetic field, the torsional Hall viscosity has the functional dependence $\zeta_{\rm H} = \zeta_{\rm H}(B^2 - \vec{E}^2)$. Thus, upon replacing $B^2\rightarrow -\vec{E}^2$ in the asymptotic series in Eq.~\eqref{Eq:TorZetaPertB}, we obtain the asymptotic series for $\zeta_{\rm H}$ in powers of the electric field,
\begin{equation}
        \label{Eq:TorHallPertInE}
        \begin{aligned}
        &\zeta_{\rm H}(\Vec{E})\sim {m^2 \over 8\pi^{3/2}}\sum_{k=0}^\infty {(-1)^k B_{2k}\Gamma(2k-1/2)\over \Gamma(2k+1)}\left({2e|E| \over m^2} \right)^{2k}\,,
        \end{aligned}
\end{equation}
with $|E| = \sqrt{\vec{E}^2}$. 
{{The above asymptotic series for $\zeta_{\rm H}$ has an important feature: In contrast to the magnetic field case, Eq.~\eqref{Eq:TorZetaPertB}, the series in Eq.~\eqref{Eq:TorHallPertInE} is non-alternating. This implies that its inverse Borel transform contains poles along the resummation contour, and, hence, the Borel resummation is not defined in a unique way. That is, we cannot resum Eq.~\eqref{Eq:TorHallPertInE}  by simply replacing $B^2\rightarrow -\vec{E}^2$ in Eq.~\eqref{Eq:TorViscInB}. Physically, this is a consequence of the electric field driving the Dirac fermion out-of-equilibrium and not allowing it to settle in a stable vacuum state. Hence, the series in Eq.~\eqref{Eq:TorHallPertInE} must be treated carefully since it must contain information from the non-perturbative sectors of the theory. These non-perturbative and non-equilibrium contributions can be evaluated via the techniques of resurgence, as described in Refs.~\cite{DUNNE_2005,Aniceto_2019}. }}

\begin{figure}[h]
    \centering
    \includegraphics[scale=0.25]{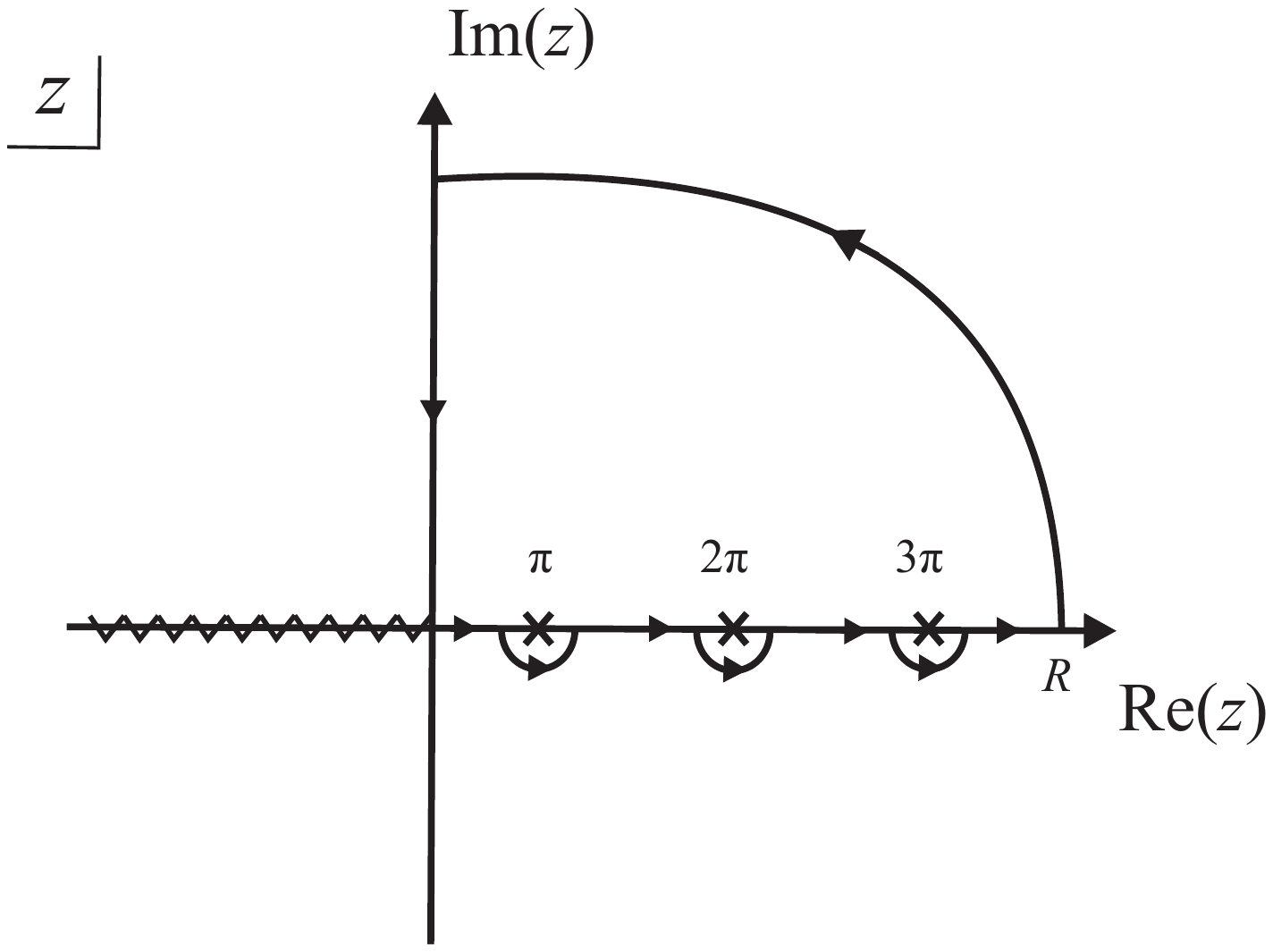}
    \caption{Contour used for the evaluation of $I(x)$ in Eq.~\eqref{Eq:AuxIntegral}. Taking $R\to\infty$ we pick up the rest of the poles of the integrand.}
    \label{fig:contour}
\end{figure}

To resum the perturbation series in Eq.~\eqref{Eq:TorHallPertInE}, we rewrite it as
\begin{align}
    \zeta_{\rm H} 
     &= \sgn(m){m^2 \over 4\pi}\left[1 -{1\over 2\sqrt{\pi}}S\left({|eE|\over \pi m^2}\right)\right]~,
\end{align}
with 
\begin{equation}
   S(x) = \sum_{k=1}^{\infty }{(-1)^kB_{2k}\Gamma(2k-1/2)\over \Gamma(2k+1)}(2\pi x)^{2k}
\end{equation}
being a formally divergent sum. To resum this expression, we evaluate it as follows:
\begin{align}
    S(x) 
    &=-2\sum_{k=1}^{\infty }\zeta(2k)\Gamma(2k-1/2)x^{2k}\nn
    \\
    &=-2\int_0^\infty dt~{\rm e}^{-t}\sum_{k=1}^{\infty }t^{2k-1/2-1}\zeta(2k)x^{2k}\nn
    \\
    &=-2\int_0^\infty dt~{{\rm e}^{-t}\over t^{3/2}}\left[\sum_{k=0}^{\infty }\zeta(2k)(tx)^{2k}- \zeta(0)\right]\nn
    \\
    &=\int_0^\infty dt~{{\rm e}^{-t}\over t^{3/2}}(\pi tx \cot(\pi tx)-1)\nn\\
    &\equiv \pi I(x)-\Gamma(-1/2)= \pi I(x) + 2\sqrt{\pi}\,,
\end{align}
where in the first equality we expressed the Bernoulli numbers in terms of the $\zeta$-function, and in the second equality we used the integral representation of the $\Gamma$-function. We then identified the resulting sum as the generating function of the $\zeta$-function on even integers and expressed $S(x)$ as a constant plus the integral $I(x)$. This integral is defined as
\begin{align}
\label{Eq:AuxIntegral}
    I(x) = x\int_0^\infty dt~{{\rm e}^{-t}\over t^{1/2}} \cot(\pi tx) = \sqrt{x}\int_0^\infty dt~{{\rm e}^{-t/x}\over t^{1/2}} \cot(\pi t)\,.
\end{align}
Although the above integral is formally divergent, we can use complex analysis to evaluate it. The integrand of $I(x)$ (for $x>0$) has a branch cut along the negative real axis and simple poles at $t = n\geq 1\in \mathbb{Z}$. To evaluate $I(x)$, we choose the contour shown in Fig.~\ref{fig:contour}, which encircles the poles in a counterclockwise fashion. As $R\to\infty$ we pick up the residues of the integrand at the poles, 
\begin{equation}
{\rm Res}(t \rightarrow n) = {{\rm e}^{-n/x}\over \pi\sqrt{n}}~,
\end{equation}
and we may express $I(x)$ as 
\begin{equation}
    I(x) = 2\pi i\sqrt{x}\sum {\rm Res} + \sqrt{ix}\int_0^\infty dt~{e^{-it/x}\over \sqrt{t}}\coth(\pi t)~.
\end{equation}
We recognize the second integral as $I(\Tilde{m}^2)$ of Eq.~\eqref{Eq:AuxIm}, now evaluated at a purely imaginary mass. While this integral is oscillatory, we assume that it is simply the analytic continuation of $I(\Tilde{m}^2)$ at complex values and write
\begin{equation}
    \int_0^\infty dt~{e^{-it/x}\over \sqrt{t}}\coth(\pi t) = {1\over \sqrt{2}}\left[2 \zeta(1/2, 1/2\pi i x)- \sqrt{2\pi i x}\right]~.
\end{equation}
Collecting everything together, we are led to the final answer for the resummation of $\zeta_{\rm H}$. As expected, the final answer contains the answer stemming directly from Lorentz invariance and the substitution $B^2\rightarrow -\vec{E}^2$ in the magnetic field result in Eq.~\eqref{Eq:TorViscInB}. However, it also contains explicitly non-perturbative contributions which are physically present due to particle creation effects in a constant electric field. Namely, 
\begin{align}
\label{Eq:TorViscinE}
    \zeta_{\rm H}=&-{m^2v_{\rm F}^4 \over 16\pi}\sqrt{2i |E|\over E_c}\left[2 \zeta\left({1\over 2},{-iE_c\over 2|E|}\right) - \sqrt{i2|E|\over E_c}\right]
    \nonumber
    \\
    & +i {m^2v_{\rm F}^4 \over 4}\sqrt{|E|\over \pi E_c}{\rm Li}_{1/2}\left({\rm e}^{-\pi E_c/|E|}\right)~.
\end{align}
We have reverted to SI units and introduced the Schwinger electric field $E_c = m^2v_{\rm F}^3/|e|\hbar$. We also identified the infinite sum over residues with the polylogarithm ${\rm Li}_s(z)$ for $s=1/2$.

Mathematically, what our analysis has shown (with a physicist's level of rigor) is the existence of a Stokes line for $\zeta(s,a)$ at ${\rm arg}(a) = \pi/2$, where a collection of exponentially suppressed terms are turned on. By symmetry, we also expect an anti-Stokes line at $\arg(a)=-\pi/2$. These comprise the full list of Stokes and anti-Stokes lines for the $\zeta$-function, as was rigorously proved in Ref.~\cite{Paris2005}. 

\begin{figure}[h]
    \centering
    \includegraphics[scale=0.75]{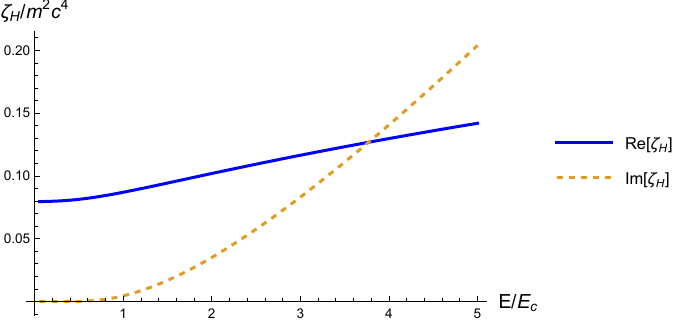}
    \caption{Plot of the real (blue/continuous curve) and imaginary (orange/dashed curve) parts of $\zeta_{\rm H}$ in units of the electron rest mass as a function of the dimensionless combination $|E|/E_c$.}
    \label{fig:ZetaR}
\end{figure}

In Fig.~\ref{fig:ZetaR}, we plot $\zeta_{\rm H}$ as a function of the electric field in dimensionless units. We observe that the torsional Hall viscosity is largely independent of the electric field until we reach $|E|\sim E_c$. After that threshold, $\zeta_{\rm H}$ develops a weak dependence on the electric field, along with a nontrivial imaginary part. {{We believe the emergence of the imaginary part is an indication of a vacuum instability due to electron-hole pair creation and the Schwinger effect \cite{PhysRev.82.664}. Hence, our results are not applicable to hydrodynamics\footnote{A prerequisite of hydrodynamics is to put the system in a thermal state, which is impossible without relaxing the energy added by the electric field \cite{Amoretti:2022ovc,Amoretti:2024jig}.}---though they can be relevant for spin transport beyond the hydrodynamic regime. For example, non-perturbative electric fields have already been shown to affect the supercurrent in Josephson junctions (see, e.g., Ref.~\cite{Amoretti:2023dgb} and references therein)}}.

Finally, we extend our discussion to the functional form of the torsional Hall viscosity in a constant electric \textit{and} magnetic field. The following discussion presumes that the electromagnetic field is inherently $2+1$ dimensional. Otherwise, for higher dimensions, additional Lorentz scalars exist, which invalidate our argument. With this caveat in mind, we note that in general $\zeta_{\rm H} = \zeta_{\rm H}(F^2)$.  If $F^2>0$, and the magnetic field dominates, we can follow the renormalization approach used to arrive at the magnetic field result, and $\zeta_{\rm H}$ is formally given by Eq.~\eqref{Eq:TorViscInB} after substituting $|B|\rightarrow \sqrt{F^2}$. In contrast, when $F^2<0$, and the electric field dominates, the non-perturbative contributions found in Eq.~\eqref{Eq:TorViscinE} must also be included, this time after substituting $|E|\rightarrow \sqrt{-F^2}$. In this way, we see that including an electric field pushes the root of $\zeta_{\rm H}$ to higher magnetic field strengths, while a magnetic field helps to stabilize the system.

\section{Turning on the BHZ deformation}\label{sec:BHZ}    

\begin{figure}[h]
    \centering
    \subfigure[$m=0$, $b=0$]{\includegraphics[scale=0.44]{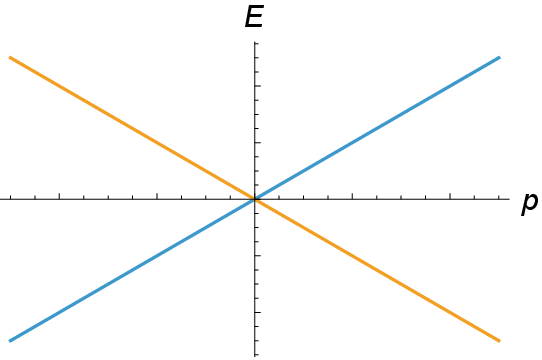}}\quad
    \subfigure[$m\neq0$, $b=0$]{\includegraphics[scale=0.44]{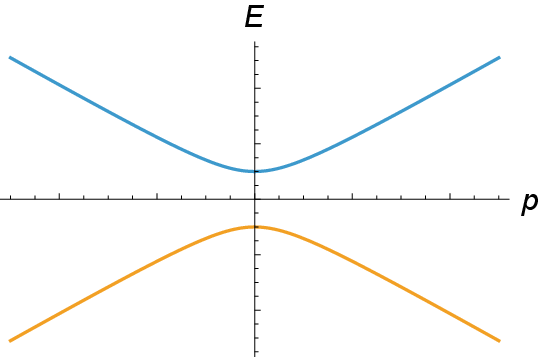}}\\
    \subfigure[$m=0$, $b\neq0$]{\includegraphics[scale=0.44]{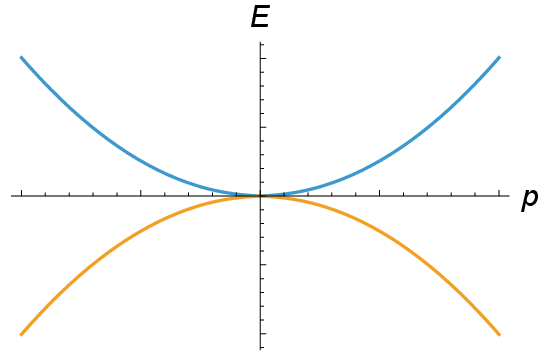}}\quad
    \subfigure[$m\neq0$, $b\neq0$]{\includegraphics[scale=0.44]{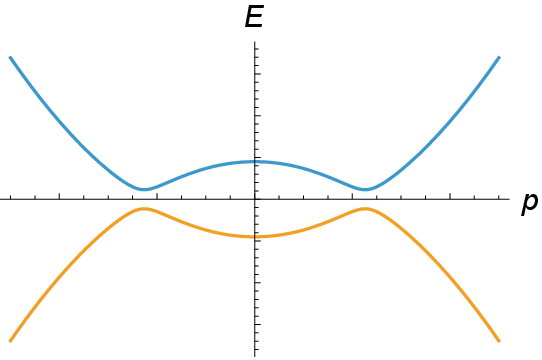}}
    \caption{The energy $E$ of the BHZ model as a function of the spatial momentum $p$ for different cases of $m$ and $b$.}
    \label{fig:energy spectrum}
\end{figure}

We now consider another important deformation of the massive Dirac fermion, quadratic in momentum, first considered in a condensed matter context for the description of 2D time reversal symmetric topological insulators (TIs) by Bernevig-Hughes-Zhang (BHZ) \cite{Bernevig_2006}. In contrast, we consider a single spin block of the BHZ Hamiltonian, which breaks parity and time-reversal symmetry:
\begin{align}
\label{HBHZ}
H=\left(\begin{array}{cc}m-bp\bar p & ap \\a\bar{p} & -m+b p\bar{p}\end{array}\right)\,,
\end{align}
where $p=p_x+ip_y$. The mass $m$ is, as before, the Dirac mass of the system. The additional parameters $a$ and $b$, for example in HgTe quantum wells, have the following physical meaning: $a$ is a measure of the hybridization between electron-like and hole-like bands, which is represented by the Fermi velocity $v_{\rm F}$, while $b$ describes the Newtonian mass of the system, i.e. the mean curvature of the bands. For $a=1$, the Lagrangian corresponding to $H$ is that of a massive relativistic Dirac fermion with a Lorentz-violating term, 
\begin{align}
\label{Eq:BHZLag}
L=\overline\psi\slashed\p\psi-m\overline\psi\psi-b\p_i\overline\psi\p^i\psi\,,
\end{align}
where $b$ has units of inverse mass in a unit system with $v_{\rm F}=\hbar=1$. In a non-relativistic setting we may identify $b =1/(2M)$, with $M$ the Newtonian mass of the fermion. To illustrate the effect of $b$ on the system and its interplay with $m$, we present the spectrum corresponding to $L$ for various values of $b$ and $m$ in Fig.~\ref{fig:energy spectrum}.

The Lagrangian in Eq.~\eqref{Eq:BHZLag} has been explored in the presence of a $U(1)$ background field \cite{Lu_2010,Bottcher:2019rrz,Tutschku:2020drw}. It was found that the Lorentz-violating term contributes to the Hall conductivity as
\begin{align}
\label{BHZconduct}
\sigma_{\rm H}=\frac{e^2}{4\pi}(\sgn(m)+\sgn(b))\,,
\end{align}
showcasing the existence of one topologically trivial and one topologically nontrivial phase depending on the relative sign of $m$ and $b$.

We shall now investigate whether and how the $b$ deformation also affects the torsional Hall viscosity $\zeta_{\rm H}$. 

\subsection{Torsional Hall viscosity from tadpole}

Since the effective action of the system is written in the form of Eq.~\eqref{Eq:ZetaEfAction}, the coefficient $\zeta_{\rm H}$ can be identified as the torsional Hall viscosity, analogous to the coefficient $\sigma_{\rm H}$ of the Chern-Simons action being the Hall conductivity. This definition relies only on the geometric structure of the background (vielbein and torsion) and does not require microscopic Lorentz invariance. Therefore, we expect that the Lorentz-violating deformation due to the BHZ term merely modifies the value of $\zeta_{\rm H}$ and does not introduce a new response coefficient at the same derivative order. Hence, the torsional Hall viscosity is given in terms of the torsion-free fermion propagator $G_{0}$ and the torsion-fermion vertex $V$, see Eqs.~\eqref{Eq:ZetaDeformedGreen} and \eqref{ZetaTadpole} in Section \ref{Sec:Setup}. 

The propagator for the BHZ Lagrangian is readily found from Eq.~\eqref{Eq:BHZLag} to be
\begin{align}
G_{0}&={1 \over \slashed{\partial} - (m + b\partial_\perp^2)}
={\slashed{\partial} + m + b\partial_\perp^2 \over -\partial^2 + (m + b \partial_\perp^2)^2}\,,
\end{align}
with $\partial_\perp^2=\partial_1^2+\partial_2^2$.

To find the vertex $V$, we must specify how the BHZ deformation term couples to the background geometry. The relativistic Dirac part admits a local $SO(3)$ spin connection $\omega^a{}_b{}_{\mu}$ as in Eq.\,\eqref{Eq:DiracActionEC}, whereas the BHZ deformation preserves only an $SO(2)$ subgroup. Thus, the BHZ term can only couple to the $\omega^1{}_2{}_{i}$ component of the spin connection. There are two possible ways forward: First, we may couple the BHZ deformation to the background geometry by replacing the spatial partial derivatives with $SO(2)$-covariant derivatives with respect to $\omega^1{}_2{}_{i}$. Second, we can treat $b$ as a ``spurion'', i.e. as a parameter which when turned on explicitly breaks the $SO(3)$ symmetry down to $SO(2)$, and allow torsion to couple to $\psi$ only through the Dirac operator in Eq.~\eqref{Eq:DiracActionEC}. We can show that both approaches are equivalent for defining $\zeta_{\rm H}$, by expanding the fermion effective action up to linear order in the connections. In what follows we adopt the minimal spurion approach and use the Lagrangian
\footnote{Including spin connections also in the BHZ term would generate extra geometric couplings that contribute only at higher order in derivatives to the stress tensor and spin current. By choosing the minimal spurion scheme we neglect these subleading effects.}
\begin{equation}
\label{Eq:TorsionBHZLag}
    L = \bar{\psi}(\slashed{\nabla}-m)\psi + b \,\partial_i\bar{\psi}\,\partial^i\psi\,.
\end{equation}
As a result, our vertex factor remains unchanged $V = 1/4$. As a consistency check of our spurion approach, we calculate $\zeta_{\rm H}$ using the tadpole diagram in Eq.~\eqref{ZetaTadpole} and compare the result to a related Berry curvature calculation in Section~\ref{subsec:BerryC}.

To evaluate the tadpole diagram, we now go to momentum space to find 
\begin{align}
\label{trGint}
\tr[G_{0}]
&=\int_{-\infty}^\infty \frac{dk_0}{2\pi}\int_0^{\infty}\frac{k_\perp dk_\perp}{2\pi}\frac{m-b k_\perp^2}{-k_0^2+ k_\perp^2+(m-b k_\perp^2)^2}\nn\\
&=-\frac{i}{4\pi}\int_0^{\infty} dk_\perp\frac{k_\perp(m-b k_\perp^2)}{\sqrt{ k_\perp^2+(m-b k_\perp^2)^2}}
\nn\\
& = -{i\over 8\pi}\int_{0}^{\infty}dy {m-by\over \sqrt{y + (m-by)^2}}\,.
\end{align}
By introducing a high-momentum cutoff $\Lambda$, we obtain (see Appendix~\ref{App:inttrG}) 
\begin{align}
\label{kperpint}
\tr[G_{0}]={}&\frac{i}{8\pi}\bigg(\sgn(b)\Big(\Lambda^2-\frac{\ln(4b^2\Lambda^2)}{2b^2}+\frac{1}{2b^2}\Big) \nonumber
\\
&-\frac{\ln(1-4bm)}{4b^2}(\sgn(m)-\sgn(b))
\nonumber\\
&-\frac{m}{b}(\sgn(b)+\sgn(m))\bigg)\,. 
\end{align}
We observe that the tadpole exhibits a quadratic and a logarithmic divergence. These divergences are encoded entirely by Dirac mass-independent, but $b$-dependent terms. This is in contrast to the massive Dirac fermion result in Eq.~\eqref{Eq:VacuumZeta1}, which exhibits a linear divergence dependent on the mass of the fermion. This stems from the fact that the BHZ term is a higher-derivative deformation and hence controls the UV behavior of the effective mass $m-bk_\perp^2$ of $\psi$. Therefore, $b$ dictates the large-momentum behavior of the tadpole integral in Eq.~\eqref{trGint}, i.e., it dictates the UV divergences we encounter. Consequently, we must examine carefully the behavior of $\zeta_{\rm H}$ as $b\rightarrow 0$ (or more precisely when $bm\ll 1$). In this limit, the large-momentum behavior of the integrand in Eq.~\eqref{trGint} is determined by the Dirac mass, plus subleading terms in $bm$. This implies that, in order to renormalize $\zeta_{\rm H}$ for the entire range of $b$, we must also renormalize an emergent linear divergence, dependent on $m$, as $b$ tends to zero. 

To understand the manifestation of the linear divergence, we must recall the physical status of the BHZ Lagrangian in Eq.~\eqref{Eq:BHZLag} as an effective field theory (EFT) of a more complicated band structure near a Dirac point. From an EFT perspective, restricting the region of the Brillouin zone around the Dirac point will reduce the importance of the BHZ deformation on $\zeta_{\rm H}$. To make this statement precise, let us assume that the BHZ deformation appears at scales of order the cutoff $\Lambda$, and re-express the BHZ Lagrangian as
\begin{equation}
\label{Eq:CutoffLag}
    L = \bar{\psi}\slashed{\partial}\psi - m \bar{\psi}\psi - \frac{\beta}{\Lambda}\,\partial_i\bar{\psi}\partial^i\psi\,,\qquad \beta \equiv b\Lambda\,.
\end{equation}
For low-energy modes with $|k_\perp|\equiv\mu\ll\Lambda$, the quadratic term is suppressed by $1/\Lambda$, and its relative size compared to the Dirac term scales as $(\beta/\Lambda)\mu^2/\mu \sim \beta\mu/\Lambda \ll 1$. Thus, in the IR the BHZ term is an irrelevant deformation and relativistic dynamics is recovered. However, since the BHZ deformation controls the UV tail of the effective mass $m-bk^2$, it contributes to the action a \emph{dangerously irrelevant} operator, which fixes the quadratic and logarithmic UV behavior of the theory.

The EFT perspective clarifies how we must proceed in order to renormalize $\zeta_{\rm H}$ for both finite-$b$ and in the $b\to0$ limit. At fixed $b\neq 0$, the UV is controlled by $bk^2$. The tadpole therefore exhibits quadratic and logarithmic divergences that are $m$-independent but $b$-dependent, see Eq.~\eqref{kperpint}. These divergences are removed by local counterterms (or Pauli-Villars fields) tailored to the $b$-dominated UV behavior, leaving a finite $\zeta_{\rm H}$. On the other hand if we take $b\to0$ and $\Lambda\to \infty$, while keeping $\beta= b\Lambda = \mathcal{O}(1)$, the UV reverts to a Dirac behavior and we recover the linear in $m$ divergence observed in Eq.\,\eqref{Eq:VacuumZeta1}. 
Therefore, to obtain a well-defined Dirac limit, we must introduce additional $b$-independent counterterms into our renormalization scheme to remove the $m$-dependent divergence. Practically, this implies that we must impose an additional renormalization condition for the Dirac mass.

With this discussion in mind, we introduce a Pauli-Villars regulator to renormalize Eq.~\eqref{kperpint}. In particular, we introduce $N$ Pauli-Villars fields with couplings $m_i$ and $b_i$, and weights $c_i$, $i=1,\cdots,N$. To cancel the quadratic and logarithmic divergences, we require 
\begin{align}
\label{PVreg1}
&\sum_{i=0}^N c_i\sgn(b_i)=0\,,\\
\label{PVreg2}
&\sum_{i=0}^N c_i\frac{\sgn(b_i)}{b_i^2}=0\,,\\
\label{PVreg3}
&\sum_{i=0}^N c_i\frac{\sgn(b_i)}{b_i^2}\ln(|b_i|)=0\,,
\end{align}
where $i= 0$ denotes the parameters of the original fermion, e.g., $b_0=b$. To obey Eq.~\eqref{PVreg1}, we can take $N=3$ and set
\begin{align}
\label{PVreg1c}
&c_0=1\,,\quad c_1=-1\,,\quad c_2=-1\,,\quad c_3=1\,,\\
\label{PVreg1b}
&\sgn(b_1)=\sgn(b_2)=\sgn(b_3)=\sgn(b)\,.
\end{align}
The condition in Eq.~\eqref{PVreg2} then allows us to set
\begin{align}
\label{PVreg2b}
&\frac{1}{b_1^2}=\frac{1}{b^2}+\frac{1}{\epsilon_1^2}\,,\qquad
\frac{1}{b_2^2}=\frac{1}{b^2}+\frac{1}{\epsilon_2^2}\,,\\
&\frac{1}{b_3^2}=\frac{1}{b^2}+\frac{1}{\epsilon_1^2}+\frac{1}{\epsilon_2^2}\,,
\end{align}
with $1/\epsilon_i\to \infty$. Equivalently, we can express the renormalization constraints in terms of the corresponding Newtonian masses $2M_i=1/b_i$ $(2M_0=2M=1/b)$ as
\begin{align}
&\sgn(M_1)=\sgn(M_2)=-\sgn(M_3)=-\sgn(M)\,,\\
&M_1^2=M^2+\tilde M_1^2\,,\qquad
M_2^2=M^2+\tilde M_2^2\,,\nn
\\
&M_3^2=M^2+\tilde M_1^2+\tilde M_2^2\,,
\end{align}
where $\tilde M_i=1/(2\epsilon_i)$, $i=1,2$. 
The condition in Eq.~\eqref{PVreg3} places a further constraint on $\tilde M_1$ and $\tilde M_2$, making them non-independent.
\par
Upon removing the Newtonian mass contribution of the Pauli-Villars regulators, we assume $b_i\Lambda = {\cal O}(1)$, in order to gain access to the ``low-energy'' divergences due to their corresponding Dirac masses $m_i$. As elaborated in Ref.~\cite{Hughes2013}, the renormalization conditions for the relativistic masses are\footnote{The full set of conditions for $m_i$ stems also from regularizing the thermal Hall coefficient $\kappa_{\rm H}$ and the Hall conductivity $\sigma_{\rm H}$.}
\begin{align}
\label{mPVcondition}
\sum_{i=0}^3c_i=0\,,\qquad\sum_{i=0}^3c_im_i=0\,,\qquad\sum_{i=0}^3c_im_i^2=0\,.
\end{align}
The first condition is already satisfied by Eq.~\eqref{PVreg1c}. For simplicity we also take $\sgn(m_1)=\sgn(m_2)=\sgn(m_3)=1$.

After properly renormalizing both $M_i$ and $m_i$, the universal part of the result gives the torsional Hall viscosity (see Appendix~\ref{App:inttrG}):
\begin{widetext}
\begin{align}
\label{Eq:ZetaBHZNonuni}
\zeta_{\rm H}={}&\frac{1}{4\pi}\bigg(mM(\sgn(m)+\sgn(M))+\frac{M^2}{2}\ln\Big(1-\frac{2m}{M}\Big)(\sgn(m)-\sgn(M))+m^2(1-\sgn(M))\bigg)\nn\\
={}&\frac{1}{4\pi}\bigg(\frac{m}{2b}(\sgn(m)+\sgn(b))+\frac{1}{8b^2}\ln\Big(1-4bm\Big)(\sgn(m)-\sgn(b))+m^2(1-\sgn(b))\bigg)\,.
\end{align}
\end{widetext}
We can see that this expression contains three parts. The first two parts come from the finite part of Eq.~\eqref{kperpint}, which depends only on the relative sign of $b$ and $m$. The remaining part is a residue of the renormalization procedure and depends solely on the sign of $b$. 

This situation is similar to the massive Dirac fermion result. In that case, the finite part of $\zeta_{\rm H}$ depends solely on the sign of $m$, but renormalization generates a term agnostic to it. In the Dirac fermion case, this latter term chooses between the $m > 0$ or the $m<0$ phase being the trivial phase of the system, where the value for $\zeta_{\rm H}$ vanishes. In the $b\neq0$ case we have no phase with vanishing $\zeta_{\rm H}$, but something similar happens: while prior to renormalization the phases depend only on the sign of $bm$, renormalization forces an explicit dependence on both $m$ and $b$. Thus, upon renormalization, we end up with four distinct phases. Although $\zeta_{\rm H}$ is nonvanishing in all these phases, we recognize the cases with $\sgn(bm)>0$ to be topological and $\sgn(bm)<0$ to be trivial. This characterization is inferred from Ref.~\cite{Tutschku:2020drw} and the result for the BHZ Hall conductivity. We also note that our choice of renormalization procedure renders $\zeta_{\rm H}$ non-covariant with respect to the time-reversal operator, which acts by flipping the sign of both $m$ and $b$. This does not lead to any physical inconsistency, since only the difference in values between different phases is scheme independent. This can be observed from the regulated values for $\zeta_{\rm H}$ at zero and non-zero $b$, see Eqs.~\eqref{Eq:VacuumZeta1} and \eqref{kperpint} respectively.

Given a fixed sign for $b$, the sign of $m$ distinguishes two distinct phases:
\begin{align}
\label{zetab>0}
b>0:\quad \zeta_{\rm H}&=
\begin{cases}
\frac{m}{4\pi b}\,,& m>0\,,\\
-\frac{1}{16\pi b^2}\ln(1-4bm)\,,&m<0\,,
\end{cases}
\\
\label{zetab<0}
b<0:\quad \zeta_{\rm H}&=
\begin{cases}
\frac{1}{16\pi b^2}\ln(1-4bm)+\frac{m^2}{2\pi}\,,&m>0\,,\\
-\frac{m}{4\pi b}+\frac{m^2}{2\pi}\,,& m<0\,.
\end{cases}
\end{align}
For a given condensed matter system, the sign of $b$ is fixed by the curvature of the bands, while the sign of $m$ is related to the topology. For example, for QAH insulators such as (Hg,Mn)Te QWs, $m$ is related to the quantum well width and the concentration of magnetic impurities (manganese) \cite{Shamim2022}. Hence, the distinction between phases shown in the above equations provides us with the following physical picture: We can construct a system with a fixed sign of $b$ and a domain wall geometry, separating two samples with opposite signs of $m$. Then, in each case the domain wall separates two different phases with $\zeta_{\rm H}$ given either by $\sim m/b$ or $\sim\ln(1-4bm)/b^2$. These stem from the vacuum sector of the theory and are, hence, renormalization-scheme independent. The jump of $\zeta_{\rm H}$ across the domain wall provides the anomaly inflow for the domain wall fermion theory with torsional anomaly, as discussed in Ref.~\cite{Hughes2013}.

Our results are consistent with previous results reported in Ref.~\cite{Hughes2013} in the following sense. Upon taking $b\to 0$, the $b$ dependence gives rise to a large cutoff $\Lambda$ across the domain wall. Under renormalization, the phase with positive $m$ becomes trivial and the phase with negative $m$ remains nontrivial (see Appendix~\ref{App:inttrG}), which recovers the results for the Dirac fermion at $b=0$. The values of $\zeta_{\rm H}$ for each phase as well as their $b\rightarrow 0$ limits are listed in Table~\ref{table1}. This is the second main result of this paper.

\begin{table}[!h]
\centering
\caption{Values of $\zeta_{\rm H}$ in different phases and their limits as $b\to0$. The values at $b=0$ are not the naive limit but are renormalized again.}
\begin{tabular}{c|c|c|c}
\toprule
$\sgn(m)$&$\sgn(b)$& $4\pi\zeta_{\rm H}$ &$4\pi\zeta_{\rm H}$ $(b\to0)$\\
\midrule
$+$&$+$&$\frac{m}{b}$&$0$\\
$+$&$-$&$\frac{1}{4b^2}\ln(1-4bm)+2m^2$&$0$\\
$-$&$+$&$-\frac{1}{4b^2}\ln(1-4bm)$&$2m^2$\\
$-$&$-$&$-\frac{m}{b}+2m^2$&$2m^2$\\
\bottomrule
\end{tabular}
\label{table1}
\end{table}

For a concrete application of our results, we may realize half of the BHZ model in a QAH such as a (Hg,Mn)Te quantum well, where\footnote{Note that we follow the convention that the mass has $v_{\rm F}^2$ absorbed and hence has the dimension of energy.} $m=-25$~meV, $b=-1075$~meV~nm$^2$, and $a=\hbar v_{\rm F}=365$ meV~nm \cite{Bottcher:2019rrz}. Given these values, we can estimate how strongly the BHZ deformation affects $\zeta_{\rm H}$ compared to its value at $b=0$. Restoring SI units in our expressions for $\zeta_{\rm H}$ we have for $b<0$ and $m<0$,
\begin{align}
\zeta_{\rm H}=-\frac{m\hbar}{4\pi b}+\frac{m^2}{2\pi\hbar v_{\rm F}^2}\,,
\end{align}
while for $b=0$ and $m<0$,
\begin{align}
\zeta_{\rm H}^0=\frac{m^2}{2\pi\hbar v_{\rm F}^2}\,.
\end{align}
The ratio of the above two cases reads
\begin{align}
\label{zetaratio}
\frac{\zeta_{\rm H}}{\zeta_{\rm H}^0}=1-\frac{\hbar^2 v_{\rm F}^2}{2mb}=1-\frac{a^2}{2mb}\,,
\end{align}
which for the values of $a$, $m$, and $b$ mentioned earlier becomes $\zeta_{\rm H}/\zeta_{\rm H}^0=-1.48$. Thus, the $\zeta_{\rm H}$ in a (Hg,Mn)Te QAH insulator is negative and its magnitude is enhanced due to the BHZ deformation. We stress that this absolute bulk value is regularization-scheme dependent, similarly to Chern-Simons-type response in the pure Dirac case. Therefore, the estimate according to Eq.\eqref{zetaratio} should be viewed as an illustrative example rather than a prediction for real materials.

To arrive at a scheme-independent prediction, we consider the domain wall configuration mentioned above. For any fixed value of $b$, take $m=+|m|$ on one side of the domain wall and $m=-|m|$ on the other. As can be seen from Eqs.~\eqref{zetab>0} and \eqref{zetab<0}, the difference of the torsional Hall viscosity across this interface is a scheme-independent universal quantity:
\begin{align}
\Delta\zeta_{\rm H}
&=\frac{|m/b|\,\hbar}{4\pi}
+\frac{\hbar^3 v_{\rm F}^2}{16\pi b^2}
\ln\Big(1+\frac{4|bm|}{\hbar^2 v_{\rm F}^2}\Big)\,.
\end{align}
Comparing this with the corresponding result for $b=0$, we find that their ratio is
\begin{align}
\frac{\Delta\zeta_{\rm H}}
{\Delta\zeta_{\rm H}^{0}}
=
\frac{a^2}{2|bm|}
+
\frac{a^4}{8b^2m^2}
\ln\left(
1+\frac{4|bm|}{a^2}
\right)\,.
\end{align}
For the case of (Hg,Mn)Te quantum wells considered above, this gives $\Delta\zeta_{\rm H}/\Delta\zeta_{\rm H}^0={4.30}$. Thus, at such a domain wall the universal jump in the torsional Hall viscosity is enhanced compared to the pure Dirac fermion case. This jump is the quantity that would be probed experimentally in an interface geometry, in close analogy with how the jump in the Hall conductivity is measured across a topological phase boundary.

\subsection{Berry curvature}\label{subsec:BerryC}
Originally, the Hall viscosity for nonrelativistic fermions was derived in Ref.~\cite{Avron:1995fg} through a Berry curvature analysis.\footnote{In band theory, the Berry phase typically refers to the phase accumulated by a free fermion state as it adiabatically moves along a closed loop in the Brillouin zone. Following Refs.~\cite{Avron:1995fg,Hughes2013}, here we consider this notion in a broader context, where the state evolves adiabatically in a more general parameter space.} Applying this method to the Dirac fermion, the calculation for torsional Hall viscosity was presented in Ref.~\cite{Hughes2013}. By closely following these derivations, we now compute the torsional Hall viscosity for the BHZ model using this method. Essentially, the Berry curvature induced by an area-preserving diffeomorphism is proportional to the torsional Hall viscosity \cite{PhysRevE.56.6173}. Therefore, we can use this method to verify that our tadpole calculation, and hence our choice of torsionful BHZ Lagrangian in Eq.~\eqref{Eq:TorsionBHZLag}, is consistent.

Consider the spacetime manifold to be $T^2\times\bb R$, where the spatial torus is made of a square in $\bb R^2$ with $(x,y)\sim(x+iL,y+jL)$, $\forall i,j\in\bb Z$. Under an area preserving diffeomorphism, we consider the spatial frame and coframe to be deformed as follows: 
\begin{align}
e^1&=\frac{1}{\sqrt{\tau_2}}(dx+\tau_1dy)\,,\qquad e^2=\sqrt{\tau_2}dy\,,\\
\label{tauframe}
\un e_1&=\sqrt{\tau_2}\un\p_x\,,\qquad \un e_2=\frac{1}{\sqrt{\tau_2}}(-\tau_1\un\p_x+\un\p_y)\,.
\end{align}
This is equivalent to transforming the flat spatial metric into the following form:
\begin{align}
g=e^1\otimes e^1+e^2\otimes e^2=\frac{1}{\tau_2}\left(\begin{array}{cc}1 & \tau_1 \\\tau_1 & |\tau|^2\end{array}\right)\,,
\end{align}
where $|\tau|^2 = \tau_1^2 + \tau_2^2$. Deforming the geometry adiabatically turns the BHZ Hamiltonian into a function of $\tau_1$ and $\tau_2$. We calculate the Berry curvature in the space of these parameters. The mass $m$ and deformation parameter $b$ are assumed to be constants as far as this calculation is concerned. 

The deformed BHZ Hamiltonian in Eq.~\eqref{HBHZ} then reads
\begin{align}
\label{Hdeformed}
H=\left(\begin{array}{cc}m-b\bf p\bar{\bf p} & \bf p \\\bar{\bf p} & -m+b\bf p\bar{\bf p}\end{array}\right)\,,
\end{align}
where ${\bf p}=(\bar\tau p_x-p_y)/\sqrt{\tau_2}$ and $\tau = \tau_1 + i \tau_2$, and we set $a =1$. The energies $E$ of the system follow as
\begin{align}
E^2={\bf p}\bar{\bf p}+(m-b{\bf p}\bar{\bf p})^2\,,
\end{align}
with eigenstates 
\begin{align}
\psi_+&=\left(\begin{array}{c}\eta\sqrt{\frac{|E|+m-b\bf p\bar{\bf p}}{2|E|}}\\\bar{\eta}\sqrt{\frac{|E|-m+b\bf p\bar{\bf p}}{2|E|}} \end{array}\right)\,,
\\
\psi_-&=\left(\begin{array}{c}\eta\sqrt{\frac{|E|-m+b\bf p\bar{\bf p}}{2|E|}}\\-\bar{\eta}\sqrt{\frac{|E|+m-b\bf p\bar{\bf p}}{2|E|}} \end{array}\right)\,,
\end{align}
where $\eta=({\bf p}/\bar{\bf p})^{1/4}$. On the torus, the spin structure is imposed by requiring the (anti)periodicity of the wavefunctions
\begin{align}
\psi(x+i,y+j)=e^{i\pi(hi+kj)}\psi(x,y)\,,~ i,j\in\bb{Z}\,.
\end{align}
where $h,k=0$ ($h,k=1$) corresponds to periodic (antiperiodic) boundary conditions. Then, momentum takes the discrete values
\begin{align}
p_x=2\pi(q+h/2)~,~ p_y=2\pi(r+k/2)~,~ q,r\in\bb Z\,.
\end{align}

Given the eigenstates, we may compute the Berry curvature. To this end, we assume that the ground state of the system is an insulator with all negative energy states occupied. Then, the Berry connection can be written as
\begin{align}
{\cal A}&=i\sum_{q,r}\langle\psi^{\dag}_-|d|\psi_-\rangle
\nonumber
\\
&=-\frac{i}{4}\sum_{q,r}\frac{m-b{\bf p}\bar{\bf p}}{\sqrt{{\bf p}\bar{\bf p}+(m-b{\bf p}\bar{\bf p})^2}}\Big(\frac{d{\bf p}}{\bf p}-\frac{d\bar{\bf p}}{\bar{\bf p}}\Big)~,
\end{align}
and its Berry curvature ${\cal F}=d{\cal A}$ reads
\begin{align}
\label{Berrycurvature}
{\cal F}=\frac{i}{8\tau_2}d\tau\wedge d\bar\tau\sum_{q,r}p_x^2\frac{m+b{\bf p}\bar{\bf p}}{({\bf p}\bar{\bf p}+(m-b{\bf p}\bar{\bf p})^2)^{3/2}}\,,
\end{align}
where we used the fact that
\begin{align}
d{\bf p}\wedge d\bar{\bf p}=-\frac{p_x^2}{2\tau_2}d\tau\wedge d\bar\tau\,.
\end{align}
In the large-volume limit, the discrete sum becomes an integral:
\begin{align}
&\sum_{q,r}p_x^2\frac{m+b{\bf p}\bar{\bf p}}{({\bf p}\bar{\bf p}+(m-b{\bf p}\bar{\bf p})^2)^{3/2}} \nonumber
\\
\to{} &\,L^2\int\frac{d^2p}{(2\pi)^2}p_x^2\frac{m+b{\bf p}\bar{\bf p}}{({\bf p}\bar{\bf p}+(m-b{\bf p}\bar{\bf p})^2)^{3/2}}\,.
\end{align}
We evaluate the integral by turning to polar coordinates, where $\sqrt{\tau_2}p_x=||p||\cos\theta$, with $||p||^2={\bf p}\bar{\bf p}$. Denoting $y \equiv ||p||^2$ for simplicity, we can now write Eq.~\eqref{Berrycurvature} as
\begin{align}
\label{BerryF}
{\cal F}&=-\frac{iL^2}{64\pi}\frac{d\tau\wedge d\bar\tau}{\tau_2^2}\int_0^\infty dy \frac{y(m+by)}{(y+(m-by)^2)^{3/2}}\,.
\end{align}
To see that the final result for ${\cal F}$ is indeed proportional to the torsional Hall viscosity, consider the remaining integral over $y$, we have 
\begin{align}
&\int_0^\infty\! dy \frac{y(m+by)}{(y+(m-by)^2)^{3/2}}=\int_0^\infty\! dyy\frac{d}{dy}\! \frac{-2(m-by)}{\sqrt{y+(m-by)^2}}\nn\\
\label{surfaceterm}
&\simeq\bigg(2\sgn(b)\Lambda^2-\frac{\sgn(b)}{b^2}\bigg)+2\int_0^\infty\! dy \frac{m-by}{\sqrt{y+(m-by)^2}}\,,
\end{align}
where in the second equality we considered a momentum cutoff $\Lambda$ and applied the large-$\Lambda$ approximation. The integral in the last line is precisely the one we used in Eq.~\eqref{trGint} to evaluate $\zeta_{\rm H}$. The accompanying surface term does not spoil this result. As we have seen, around Eq.~\eqref{kperpint}, the surface term does not contribute to the Hall viscosity, as it is removed upon renormalization.   

To recap, we have shown that the Berry curvature of the BHZ Hamiltonian on a torus is proportional to the torsional Hall viscosity we present in Eq.~\eqref{Eq:ZetaBHZNonuni}. This acts as a consistency check on our calculation and gives credence to our choice of torsionful BHZ Lagrangian.

\section{Conclusions}\label{sec:Conclusions}

We have examined how deforming the massive Dirac fermion Lagrangian alters the corresponding torsional Hall viscosity. We focused on two experimentally relevant deformations, a constant external electromagnetic field and a momentum-dependent mass term (the BHZ deformation). 

We began our discussion with a constant magnetic field deformation in Section~\ref{sec:MagField}. We derived a closed-form expression in Eq.~\eqref{Eq:TorViscInB} for $\zeta_{\rm H}$ for any value of the magnetic field. We found that $\zeta_{\rm H}$ decreases monotonically with increasing magnetic field strength, and reaches zero at $B = B_0 \neq 0$. Numerically, we found that $B_0$ appears to be roughly five times the Schwinger limit of QED, where the vacuum becomes birefringent. Using realistic parameters for Chern insulators, we found $B_0 = {\cal O}(100{\rm mT})$. These results thus offer a direct experimental probe of non-perturbative physics in condensed-matter tabletop experiments. We also provide the perturbation series to all orders in the magnetic field, which can serve as an accurate approximation when $|B|/B_c \ll 1$. The leading order correction to the free vacuum result is quadratic in $B$ with a negative coefficient. Subsequent corrections appear at even powers of $B$ ($B^4, B^6,\dots$), with alternating signs for their coefficients.

Next, we employed our results for the constant magnetic field case, as well as the Lorentz invariance of our system, to compute the torsional Hall viscosity in a constant electric field as presented in Eq.~\eqref{Eq:TorViscinE}. We observed that for electric fields below the Schwinger limit, the torsional Hall viscosity is largely independent of the electric field.  Beyond this limit, $\zeta_{\rm H}$ increases with the electric field strength and develops a nontrivial imaginary part. This imaginary part arises from non-perturbative particle-creation effects, which we interpret as the electric field breaking unitarity and destabilizing the Dirac fermion vacuum.

We then turned our attention to the BHZ model as a more realistic description of fermion transport in tabletop experiments compared to the massive Dirac fermion. This model breaks Lorentz invariance down to rotational invariance by introducing a mass deformation that depends on the spatial momentum of the fermion. By introducing Pauli-Villars regulator fields and renormalizing the tadpole diagram, we derived the expression Eq.~\eqref{Eq:ZetaBHZNonuni} for the torsional Hall viscosity of this model, which distinguishes between topologically trivial and nontrivial phases. These phases resemble the undeformed massive Dirac fermion case \cite{Hughes2013}, where the Hall viscosity is determined solely by the sign of $m$, but here the interplay of $m$ and $b$ introduces richer behavior. We validated our results through a Berry curvature analysis, which confirms that the torsional Hall viscosity arises from adiabatic deformations of the spatial metric. This is consistent with its interpretation as a geometric response.

The final result exhibits several key features. First, $\zeta_{\rm H}$ manifests both polynomial and logarithmic dependencies on $m$ and $b$. Applying our analysis to a realistic system, namely a (Hg,Mn)Te quantum well, we found that the BHZ deformation can substantially modify the bulk torsional Hall viscosity compared to the undeformed massive Dirac fermion case. For a domain wall across which the Dirac mass of the (Hg,Mn)Te quantum well changes sign, we found that the universal jump $\Delta\zeta_{\rm H}$ is enhanced by a factor of about $4.3$ compared to the undeformed Dirac case. This domain-wall jump is a scheme-independent, experimentally relevant prediction of our analysis. Strikingly, the singular $b \to 0$ limit underscores the non-perturbative nature of the BHZ deformation, as the Newtonian mass governs UV divergences that require careful renormalization to recover the corresponding result of the Dirac fermion. As shown in Table~\ref{table1}, the $b \to 0$ limit recovers the Dirac fermion result only after accounting for emergent divergences.

Our manuscript opens up several research directions. {{The first is the extension of our results to nonzero temperature and chemical potential.}} Second, it would be interesting to use the results of this paper to provide hydrodynamic simulations of $2+1$ spinful fluids in the presence of the BHZ deformation parameter or external electromagnetic fields.  Furthermore, it would be interesting to understand how momentum relaxation, included in the hydrodynamic equations \cite{Amoretti:2024jig}, interplays with the instability of the system in an electric field, as described by the nonzero imaginary part of $\zeta_{\rm H}$. {{These simulations would enable the experimental verification of our results via non-local spin measurements and the spin Hall effect, see, e.g., \cite{Brune2012}, by providing closed-form expressions of spin currents as a function of $\zeta_{\rm H}$.\footnote{A caveat: The hydrodynamic spin current will most likely depend on additional Hall coefficients and, thus, knowledge of spin-spin and spin-momentum correlations will be necessary to isolate the effects of $\zeta_{\rm H}$.} An additional way to verify our result is the following: According to the effective action in Eq.~\eqref{Eq:ZetaEfAction}, the energy-momentum current is proportional to the torsion tensor and $\zeta_{\rm H}$. Therefore, measuring the energy-momentum current of the system and fixing the form of the torsion tensor through strain-engineering \cite{10.1038/nnano.2009.191,doi:10.1126/science.1191700,doi:10.1126/sciadv.aat9488} (see also reviews \cite{Amorim_2016,C5NR07755A}) allows us to measure $\zeta_{\rm H}$ directly.}} Another interesting problem is to examine whether there is a connection between the Hall conductivity and the torsional Hall viscosity along the lines of Ref.~\cite{Hoyos_2014}. Finally, we should ascertain whether the influence of the torsional Hall viscosity is the dominant one in the spin-hydro regime. Hence, it would be interesting to derive the complete list of parity-odd, non-diffusive transport coefficients entering the spin current constitutive relations, e.g., by employing and extending the AdS/CFT correspondence \cite{Gallegos:2020otk,Erdmenger:2022nhz,Erdmenger:2023hne}.

\onecolumngrid

\begin{acknowledgments}

We thank Christian Tutschku for collaboration in the early stages of this project. E.M.H.~thanks Joel Moore and James Analytis for the insightful discussions and the support as well as the hospitality at the University of California, Berkeley. W.J.~thanks Yingfei Gu for his hospitality and support at IASTU, and Chao Xu for helpful discussions. E.M.H., J.E., and R.M.~acknowledge the support of the German Research Foundation (DFG) through the Collaborative Research Center ToCoTronics, Project-ID 258499086 — SFB 1170, as well as Germany’s Excellence Strategy through the W{\"u}rzburg-Dresden Cluster of Excellence on Complexity and Topology in Quantum Matter - ct.qmat (EXC 2147, Project-ID 390858490). The work of I.M.~is supported by the STFC consolidated grant (ST/X000583/1) `New Frontiers In Particle Physics, Cosmology And Gravity'. The work of W.J. is also supported by funding from Hong Kong’s Research Grants Council (RFS2324-4S02, CRF C7015-24GF). Z.Y.X. acknowledges support from the Berlin Quantum initiative.
\end{acknowledgments}

\clearpage

\appendix

\section{Path integral derivation of the torsional Hall viscosity}\label{appendix:pathIntegralZetaH}

In this appendix, we present the derivation of Eq.~\eqref{Eq:ZetaDeformedGreen} in the main text, linking the torsional Hall viscosity to the torsion-free Green's function.

To begin, assume that the action can be written as
\begin{equation}
    S_\psi = \int d^3{x}\, e~\bar{\psi }{D}\psi ~,
\end{equation}
where $D$ is some local differential operator, and $e = |\det(e^a_{\ \mu})|$ is the invariant volume element associated with the background metric $g_{\mu\nu} = e^a_{\ \mu} e^b_{\ \nu}\eta_{ab}$.\footnote{Note that $e$ does not assume the theory to be Lorentz invariant and therefore does not need to be modified when Lorentz-violating deformations such as the BHZ term are included.}
Then, the effective action for torsion can be derived by integrating out the fermion fields $\psi$ and $\bar{\psi}$ by the fermionic (Euclidean) path integral \cite{Peskin:1995ev}
\begin{align}
        e^{-S_{\rm eff}} =\int {\cal D}\bar{\psi}{\cal D}\psi ~e^{-S_\psi}
        =\det(D)
        =e^{\tr[ \ln (D)] }~,
\end{align}
where we used $\ln[\det(D)] = \tr[ \ln (D)]$. Now assume that we can separate the operator $D = \mathring{D} + V\sigma$ in terms of a torsion-free part $\mathring{D}$ and a part linear in the torsion pseudoscalar $\sigma$ [as shown explicitly in the main text, Eq.~\eqref{eq:torsionPseudoscalarDefinition}] with a vertex function $V$, which for the moment we leave arbitrary. 
Then, the effective action becomes
\begin{equation}
\begin{aligned}
-S_{\rm eff}
&= 
\tr [ \ln ( \mathring{D}+ V{\sigma } ) ] 
=
\tr [ \ln (\mathring{D} (1+\mathring{D}^{-1}V\sigma ) ) ]
\\
&=
\tr[\ln( \mathring{D})]+
\tr[\ln ( 1+\mathring{D}^{-1}V \sigma ) ]
= \log Z_{0} +
\tr [\ln ( 1+G_0 V\sigma )]\,,
\end{aligned}
\end{equation}
where $G_0 = \mathring{D}^{-1}$ is the torsion-free Green's function, and $\log Z_0 = \tr[\ln( \mathring{D})] $ may be thought of as the effective action of the torsion-free system which can be safely ignored in the following.\footnote{Alternatively, one can normalize $Z_0$ to unity.} 
Therefore, the effective action is given by
\begin{equation}
    -S_{\rm eff} 
    = \tr[ \ln ( 1+G_0 V\sigma )]
    = \tr\left[ \sum_{n=1}^\infty \frac{(-1)^{n-1}}{n} (G_0 V\sigma)^n \right] \,.
\end{equation}
In this paper, the torsional Hall viscosity is encoded only in the part of $S_{\rm eff}$ that is linear in the torsion pseudoscalar $\sigma$. Therefore, we may restrict ourselves to the term with $n=1$,
\begin{equation}
\label{AppEq:TraceEff}
    S_{\rm eff} = - \tr[G_0V\sigma]\,.
\end{equation}
At this point, we must clarify how to take the trace on the right-hand side of Eq.~\eqref{AppEq:TraceEff}. All  operators are operators in both real space and in spin space. Therefore, in full generality, $S_{\rm eff}$ can be written as
\begin{equation}
\begin{aligned} 
S_{\rm eff }
= -\int d^{3}x\, e~d^{3}y\, e~d^{3}z\, e~\tr[ G_{0}(x;y) \, V( x;z) \, \sigma ( z;x) ]~, 
\end{aligned}
\end{equation}
with $\tr(\, \cdot\, )$ in the integral being the remaining trace over spin degrees of freedom. To proceed, we first note that both $V$ and $\sigma$ are local, e.g., $\sigma (z;x) = \sigma(x) \delta(x-z)/e$, and independent of spin. This allows us to simplify $S_{\rm eff}$ to
\begin{equation}
    S_{\rm eff} 
    = -\int d^{3}x\, e \lim_{y\to x}\tr[ G_{0}( x;y)] \, V( x)\, \sigma(x) \,.
\end{equation}
Finally, we assume that the system is homogeneous in the absence of torsion, i.e., $G_0(x;y) = G_0(x-y)$, and that the vertex function $V(x) = V$ is constant. 
In this way, we obtain the final expression for the torsional effective action
\begin{equation}
    S_{\rm eff} 
    = 
    -\lim_{y\to x}\tr[ G_{0}(x;y) ] \, V \int d^3x\, e~\sigma(x)  \equiv - \dfrac{\zeta_{\rm H}}{2}\int d^3x \, e~\sigma(x)\, , 
 \end{equation}
from which we can read off the expression of the torsional Hall viscosity 
\begin{align}
    \label{eq:generalVertex}
    \zeta_{\rm H} = 2 \, \lim_{y\to x}\tr[ G_{0}(x;y) ]  V \, .
\end{align}
For $V = 1/4$ [cf.~Eq.~\eqref{eq:torsionPseudoscalarDefinition}], we get the result presented in Eq.~\eqref{Eq:ZetaGreen} of the main text.

\section{Dirac equation solutions for finite magnetic field and vacuum heat kernel}\label{App:VacuumSol}

In this appendix, we present the solution of the Dirac equation in a constant magnetic field and calculate the corresponding heat kernel. 

The Dirac Hamiltonian we wish to diagonalize reads
\begin{equation}
    H = - \slashed{D}^2 + m^2 ~~,~~ \slashed{D} = \gamma^\mu(\partial_\mu - ieA_\mu)~.
\end{equation}
As we mentioned in the main text, we shall work in Euclidean signature in order to avoid certain divergences that appear in the Lorentzian case. Thus, to proceed we choose a particular Euclidean representation of the Dirac matrices, $\gamma^\mu = (\gamma^3, \gamma^1, \gamma^2) =(\sigma_z, \sigma_x, \sigma_y)$, where $\sigma_i$ are the usual Pauli matrices. Our constant background magnetic field $B$ is generated by the gauge potential mentioned in the main text, $A_\mu = (A_3, A_1, A_2) = (0, 0,B x^1)$. Then the eigenvalue equation $H\psi = \lambda\psi$ reads
\begin{equation}
    [-\partial_1^2 -\partial_3^2 -(\partial_2-ieBx^1)^2 - eB\sigma_z +m^2]\psi = \lambda \psi~,
\end{equation}
where $\psi$ is a spinor field. Since $H$ is cyclic in $x^3$ and $x^2$, we may assume $\psi = {\rm e}^{ip_3 x^3+ip_2 x^2}\chi(x^1)$ to find
\begin{equation}
    [-\partial_1^2 + (p_2 -eBx)^2]\chi = [\lambda + eB\sigma_z - m^2-p_3^2]\chi~.
\end{equation}
Introducing the definitions $l_B^2 = 1/|eB|$ and $X = x^1/l_B + p_2l_B$, we are left with
%
%
\begin{equation}
    [-\partial_X^2 + X^2]\chi = l_B^2[\lambda + {\rm sign}(eB)\sigma_z l_B^{-2} - m^2-p_3^2]\chi~.
\end{equation}
The left-hand side is the Hamiltonian of a $1$d harmonic oscillator with spectrum $E_n = 2n +1$, $n \in \mathbb{N}_0$. The right-hand side depends on spin, which tells us that the eigenfunctions of the Dirac Laplacian are given by two sets of solutions. These solutions are
\begin{align}
    \label{Eq:SquareSolution}
    &\psi_{r, n, p_2, p_3} = N u_r e^{ip_3x^3 + ip_2x^2}e^{-X^2/2}H_n(X)
    ~~,~~
    \lambda_r =
      {2n+ 1 -{\rm sign}(eB)r\over l_B^2} + p_3^2 +m^2~,
\end{align}
with $r = \pm 1$ the spin of the solution, $N^2 = [(2\pi)^2 \sqrt{\pi}2^n n!l_B]^{-1}$ a normalization factor and 
\begin{align}
u_+ ={1\over \sqrt{2}} 
    \begin{pmatrix}
        1
        \\
        0
    \end{pmatrix}
    ~~,~~
    u_- ={1\over \sqrt{2}} 
    \begin{pmatrix}
        0
        \\
        1
    \end{pmatrix}~.
\end{align}
We observe that upon changing the sign of  $eB$, our solution remains unaltered but the eigenvalues are exchanged, i.e., $\lambda_+ \leftrightarrow \lambda_-$. This interchange of eigenvalues does not alter the calculation of the heat kernel. Thus $\psi$ gives the spinor solutions presented in the main text.

We can now use $\psi$ to evaluate the corresponding heat kernel. We recall that the heat kernel is defined as $K(s; x,y) = \langle x| \exp(-s H) |y\rangle$.
Therefore, upon taking its trace, we have\footnote{Recall that the capitalized coordinates depend on $p_2$.} 
\begin{align}
{\rm tr}[K(s;x,y)] &= \int_{-\infty}^\infty dp_3 \int_{-\infty}^{\infty} dp_2 \sum_{n=0}^{\infty}\sum_{r=\pm} {\rm e}^{-\lambda_r s}\psi^\dagger_{r,n,p_2,p_3}(y)\psi_{r,n,p_2,p_3}(x)
\nn \\
&= {e^{-m^2 s} \over 4\pi^2 \sqrt{\pi} l_B} \sum_{n = 0}^{\infty}\int_{-\infty}^{\infty} dp_3 e^{ip_3(x^3-y^3)-p_3^2 s} \int_{-\infty}^\infty dp_2 e^{i p_2 (x^2 -y^2)}(e^{\sgn (eB)s/l_B^2} + e^{-\sgn(eB)s/l_B^2})
\nn \\
&\quad \times{e^{-(2n +1)s/l_B^2}\over 2^n n!}H_n(X) H_n(Y)e^{-(X^2 +Y^2)/2} 
\nn \\
&= {e^{-m^2 s} \cosh(s/l_B^2) \over 2\pi^2 l_B\sqrt{s}} e^{-(x^3-y^3)^2/4s}\int_{-\infty}^\infty dp_2 e^{i p_2 (x^2 -y^2)}\sum_{n=0}^{\infty}{e^{-(2n +1)s/l_B^2}\over 2^n n!}H_n(X) H_n(Y)e^{-(X^2 +Y^2)/2}
\nn \\
&={e^{-m^2 s} \cosh(s/l_B^2) \over 2\pi^2 l_B \sqrt{s}} e^{-(x^3-y^3)^2/4s}\int_{-\infty}^\infty dp_2 e^{i p_2 (x^2 -y^2)}{\exp\left[ -\coth(2s/l_B^2)(X^2 + Y^2)/2 + {\rm csch}(2s/l_B^2)X Y\right]\over  \sqrt{2\sinh(2s/l_B^2)} }
\nn \\
&={e^{-m^2 s}\sqrt{\coth(s/l_B^2)}\over 4\pi^2 l_B \sqrt{s}}\int_{-\infty}^\infty dp_2 \exp\left[ -\coth(2s/l_B^2)(X^2 + Y^2)/2 + {\rm csch}(2s/l_B^2)X Y\right] .
\end{align}
In the above calculation, we first substituted the explicit form of the wavefunctions and then collected terms suggestively. We proceeded by performing the Gaussian integral over $p_3$ and the sum over $n$ (which is known in closed form \cite{MehlerKernel}). 
In the last equality, we formally took the limit $y\to x $ in order to be able to perform the remaining integral over $\tilde{p}_2$.\footnote{Technically, the limit $y\rightarrow x$ is not well-defined, since the resulting integral is divergent. We proceed formally with taking the limit and show how to treat the divergence in the main text.} 
We continue by introducing the dimensionless variables $\tilde{s} = s/l_B^2$, $\tilde{p} = pl_B$, and $\tilde{x}=x/l_B$, after which we get  
\begin{equation}
    \label{Eq:TraceK}
    {\rm tr}[K(s)] = {e^{-\tilde{m}^2 \tilde{s}}\sqrt{\coth(\tilde{s})}\over 4\pi^2 l_B^3\sqrt{\tilde{s}}}\int_{-\infty}^\infty d\tilde{p_2} e^{-(\tilde{p}_2+\tilde{x}^1)^2\tanh(\tilde{s})} ={\sqrt{\pi}\over 4\pi^2l_B^3\sqrt{\tilde{s}}}e^{-\tilde{m}^2\tilde{s}}\coth(\tilde{s})~.
\end{equation}
We have thus reached the final expression of the heat kernel, as presented in Eq.~\eqref{Eq:FiniteBHeat} of the main text. 

\section{Tadpole integral and Pauli-Villars regularization}\label{App:inttrG}

In the calculation of $\zeta_{\rm H}$ and the evaluation of $\tr [G_{0}]$, we encounter the integral
\begin{align}
\int_0^{\infty} dk_\perp\frac{k_\perp(m-b k_\perp^2)}{\sqrt{ k_\perp^2+(m-b k_\perp^2)^2}}\,.
\end{align}
First, without the bounds, the integral can be performed to be (up to a constant of integration)
\begin{align}
\label{intbk2}
\int dk_\perp\frac{k_\perp(m-b k_\perp^2)}{\sqrt{ k_\perp^2+(m-b k_\perp^2)^2}}=-\frac{1}{2b}\sqrt{k_\perp^2+(m-b k_\perp^2)^2}+\frac{1}{4b^2}\ln\Bigg(\frac{|m|-\sqrt{ k_\perp^2+(m-b k_\perp^2)^2}-bk_\perp^2}{|m|-\sqrt{ k_\perp^2+(m-b k_\perp^2)^2}+bk_\perp^2}\Bigg)\,.
\end{align}
Note that this integral is fundamentally different from the case with $b=0$, for which we have
\begin{align}
\label{intb0}
\int dk_\perp\frac{k_\perp m}{\sqrt{ k_\perp^2+m^2}}=m\sqrt{k_\perp^2+m^2}\,.
\end{align}
One cannot simply take $b\to 0$ in Eq.~\eqref{intbk2} and get the result in Eq.~\eqref{intb0}. This indicates that the integral and the limit $b\to0$ do not commute in this case. As we mention in the main text, this is because of the qualitatively different UV behavior between the $b\neq 0$ and $b=0$ cases.

Let us evaluate the right-hand side of Eq.~\eqref{intbk2} at $k_\perp =0$ and $k_\perp = \Lambda$. The behavior of the first term gives
\begin{align}
-\frac{1}{2b}\sqrt{k_\perp^2+(m-b k_\perp^2)^2}\quad\to\quad
\begin{cases}
-\frac{|m|}{2b}\,,\qquad &k_\perp=0\,,\\
-\frac{|b|}{2b}\Lambda^2-\frac{1}{4b|b|}+\frac{m}{2|b|}\,,\qquad &k_\perp=\Lambda\to\infty\,.
\end{cases}
\end{align}
Therefore,
\begin{align}
-\frac{1}{2b}\sqrt{k_\perp^2+(m-b k_\perp^2)^2}\Big|_0^\Lambda&=
-\frac{|b|}{2b}\Lambda^2-\frac{1}{4b|b|}+\frac{m}{2|b|}+\frac{|m|}{2b}\nn\\
&=-\frac{\sgn(b)\Lambda^2}{2}-\frac{\sgn(b)}{4b^2}+\frac{m}{2b}(\sgn(b)+\sgn(m))\,.
\end{align}
To evaluate the second term in Eq.~\eqref{intbk2}, we consider different cases regarding the signs of $m$ and $b$. First, for $k_\perp\to0$, we have
\begin{align}
\ln\Bigg(\frac{|m|-\sqrt{ k_\perp^2+(m-b k_\perp^2)^2}-bk_\perp^2}{|m|-\sqrt{ k_\perp^2+(m-b k_\perp^2)^2}+bk_\perp^2}\Bigg)\quad\to\quad
\begin{cases}
-\ln(1-4bm)\,,\qquad &m>0\,,\\
\ln(1-4bm)\,,\qquad &m<0\,.
\end{cases}
\end{align}
For $k_\perp=\Lambda\to\infty$, we get
\begin{align}
\ln\Bigg(\frac{|m|-\sqrt{ k_\perp^2+(m-b k_\perp^2)^2}-bk_\perp^2}{|m|-\sqrt{ k_\perp^2+(m-b k_\perp^2)^2}+bk_\perp^2}\Bigg)\,\,\,\to\,\,\,
\begin{cases}
2\ln(2b\Lambda)-\ln(1-4bm)\,, &m>0, b>0\,,\\
-2\ln(-2b\Lambda)\,,&m>0,b<0\,,\\
2\ln(2b\Lambda)\,, &m<0,b>0\,,\\
-2\ln(-2b\Lambda)+\ln(1-4bm)\,,&m<0,b<0\,.\\
\end{cases}
\end{align}
Therefore,
\begin{align}
\ln\Bigg(\frac{|m|-\sqrt{ k_\perp^2+(m-b k_\perp^2)^2}-bk_\perp^2}{|m|-\sqrt{ k_\perp^2+(m-b k_\perp^2)^2}+bk_\perp^2}\Bigg)\Bigg|_0^\Lambda
&=
\begin{cases}
2\ln(2b\Lambda)\,, &m>0, b>0\,,\\
-2\ln(-2b\Lambda)+\ln(1-4bm)\,,&m>0,b<0\,,\\
2\ln(2b\Lambda)-\ln(1-4bm)\,, &m<0,b>0\,,\\
-2\ln(-2b\Lambda)\,,&m<0,b<0\,,\\
\end{cases}\nn\\
&=2\sgn(b)\ln(2|b|\Lambda)+\frac{\sgn(m)-\sgn(b)}{2}\ln(1-4bm)\,.
\end{align}
Collecting all results, we obtain
\begin{align}
\label{Eq:ReguralizedZeta}
&\int_0^{\Lambda} dk_\perp\frac{k_\perp(m-b k_\perp^2)}{\sqrt{ k_\perp^2+(m-b k_\perp^2)^2}}\nn\\
=&-\frac{\sgn(b)\Lambda^2}{2}+\frac{\sgn(b)}{2b^2}\ln(2|b|\Lambda)-\frac{\sgn(b)}{4b^2}+\frac{m}{2b}(\sgn(b)+\sgn(m))+\frac{\ln(1-4bm)}{8b^2}(\sgn(m)-\sgn(b))\,,
\end{align}
and arrive at the result in Eq.~\eqref{kperpint} of the main text, where the remaining task is to remove $\mathcal{O}(\Lambda^2)$ and logarithmic divergences by a suitable renormalization procedure.

To renormalize our result in Eq.~\eqref{Eq:ReguralizedZeta}, we introduce $N$ Pauli-Villars fields with couplings $m_i$ and $b_i$, $i=1,\cdots,N$. At the end of the renormalization process, we must take $M_i=1/(2b_i)$ to infinity, along with the introduced cutoff $\Lambda$. The discussion in Section~\ref{sec:BHZ} around Eq.~\eqref{Eq:CutoffLag} in the main text implies that we must be careful and take this limit as $b_i\rightarrow \infty$, $\Lambda\rightarrow \infty$, but keep $|b_i|\Lambda = a_i$, where $a_i$ is an ${\cal O}(1)$ positive number. This allows us to recover the correct free massive Dirac fermion result upon taking the limit $b\to 0$ in the BHZ Lagrangian. 

To regularize the quadratic and logarithmic divergences in Eq.~\eqref{Eq:ReguralizedZeta}, we use the renormalization conditions
\begin{align}
\sum_{i=0}^N c_i\sgn(b_i)&=0\,,\qquad
\sum_{i=0}^N c_i\frac{\sgn(b_i)}{b_i^2}=0\,,\qquad
\sum_{i=0}^N c_i\frac{\sgn(b_i)}{b_i^2}\ln|b_i|=0\,.
\end{align}
By setting
\begin{align}
&c_0=1\,,\qquad c_1=-1\,,\qquad c_2=-1\,,\qquad c_3=1\,,\\
&\sgn(b_1)=\sgn(b_2)=\sgn(b_3)=\sgn(b)\,,
\end{align}
we have
\begin{align}
&\sum_{i=0}^3 \frac{M_i^2}{2}\ln\Big(1-\frac{2m_i}{M_i}\Big)(\sgn(m_i)-\sgn(M_i))\nn\\
={}&\frac{M^2}{2}\ln\Big(1-\frac{2m}{M}\Big)(\sgn(m)-\sgn(M))-\frac{M_1^2}{2}\ln\Big(1-\frac{2m_1}{M_1}\Big)(\sgn(m_1)-\sgn(M))\nn\\
&-\frac{M_2^2}{2}\ln\Big(1-\frac{2m_2}{M_2}\Big)(\sgn(m_2)-\sgn(M))+\frac{M_3^2}{2}\ln\Big(1-\frac{2m_3}{M_3}\Big)(\sgn(m_3)-\sgn(M))\,.
\end{align}
Adding up the terms above and taking $M_i\gg m_i$ for $i=1,2,3$ yields
\begin{align}
4\pi\zeta_{\rm H}={}&\sum_{i=0}^3 c_im_iM_i(\sgn(M_i)+\sgn(m_i))+\sum_{i=0}^3 c_i\frac{M_i^2}{2}\ln\Big(1-\frac{2m_i}{M_i}\Big)(\sgn(m_i)-\sgn(M_i))\nn\\
\to{}&mM(\sgn(M)+\sgn(m))+\frac{M^2}{2}\ln\Big(1-\frac{2m}{M}\Big)(\sgn(m)-\sgn(M))-2\Big(m_1\frac{\Lambda}{\alpha_1}+m_2\frac{\Lambda}{\alpha_2}-m_3 \frac{\Lambda}{\alpha_3}\Big)\sgn(M)\nn\\
&+m_1^2(\sgn(m_1)-\sgn(M))+m_2^2(\sgn(m_2)-\sgn(M))-m_3^2(\sgn(m_3)-\sgn(M))\,,
\end{align}
where in the last step we took $M_i\to\infty$ while fixing $\Lambda/M_i=\alpha_i$ to be a constant.
\par
Next, we impose the renormalization conditions in Eq.~\eqref{mPVcondition} of the main text for $m_i$ (we take $\alpha_1=\alpha_2=\alpha_3=\alpha$). We also set
\begin{align}
\sgn(m_1)=\sgn(m_2)=\sgn(m_3)=1\,.
\end{align}
Then,
\begin{align}
4\pi\zeta_{\rm H}&=-2\Big(m_1+m_2-m_3 \Big)\frac{\Lambda}{\alpha}\sgn(M)+(m_1^2+m_2^2-m_3^2)(1-\sgn(M))\\
&\to mM(\sgn(M)+\sgn(m))+\frac{M^2}{2}\ln\Big(1-\frac{2m}{M}\Big)(\sgn(m)-\sgn(M))+m^2(1-\sgn(M))-2m\frac{\Lambda}{\alpha}\sgn(M)\,.\nn
\end{align}
From the universal part we read off that
\begin{align}
\zeta_{\rm H}=\frac{1}{4\pi}\bigg(mM(\sgn(M)+\sgn(m))+\frac{M^2}{2}\ln\Big(1-\frac{2m}{M}\Big)(\sgn(m)-\sgn(M))+m^2(1-\sgn(M))\bigg)\,.
\end{align}
To go back to the result for $b=0$, we further take the limit $M\to\infty$ with $\Lambda/M=\alpha$, which gives
\begin{align}
4\pi\zeta_{\rm H}
&=2mM\sgn(M)-m^2(\sgn(m)-1)-2m\frac{\Lambda}{\alpha}\sgn(M)\nn\\
&\to m^2(\sgn(m)-1)\,.
\end{align}
Therefore, in the limit $b\to0$ we recover the result
\begin{align}
\zeta_{\rm H}
={}&\frac{1}{4\pi}m^2(\sgn(m)-1)\,.
\end{align}
 
\bibliography{refs.bib}

\end{document}